\documentclass[aps,prc,twocolumn,groupedaddress,floatfix,%
               longbibliography]{revtex4-1}
\pdfoutput=1
\usepackage{amsmath,amssymb,latexsym,array,bm}
\usepackage{graphicx}

\begin{document}

\title{Monotonic properties of the shift and penetration factors}
\author{Carl~R.~Brune}
\email{brune@ohio.edu}
\affiliation{%
  Edwards Accelerator Laboratory \\
  Department of Physics and Astronomy \\
  Ohio University, Athens, Ohio 45701, USA }
\author{Gerald~M.~Hale}
\email{ghale@lanl.gov}
\author{Mark~W.~Paris}
\email{mparis@lanl.gov}
\affiliation{%
  T-2 Nuclear \& Particle Physics, Astrophysics, \& Cosmology \\
  Theoretical Division, Los Alamos National Laboratory \\
  Los Alamos, New Mexico 87545, USA }

\date{\today}

\begin{abstract}
We study derivatives of the shift and penetration factors of collision theory
with respect to energy, angular momentum, and charge.
Definitive results for the signs of these derivatives are found for the
repulsive Coulomb case.
In particular, we find that the derivative of the shift factor with respect
to energy is positive for the repulsive Coulomb case, a long
anticipated but heretofore unproven result.
These results are closely connected to the properties of the sum of squares
of the regular and irregular Coulomb functions; we also
present investigations of this quantity.
\end{abstract}

\maketitle

\section{Introduction}

The shift and penetration factors occur in the theoretical
description of nuclear, atomic, and molecular scattering and reactions,
particularly in $R$-matrix descriptions of such processes~\cite{Lan58,Des10}.
These quantities are defined to be the real and imaginary parts of
the logarithmic radial derivative of the outgoing Coulomb function,
as given below by Eq.~(\ref{eq:l_s_p}).
They play a central role in determining how physical quantities, such
as cross sections and resonance widths,  depend upon energy, angular
momentum, and charge.

This study is motivated by a desire to understand the sign of the energy
derivative of the shift factor for the repulsive Coulomb case,
as is applicable to the study of nuclear reactions. 
This sign has important implications for the relationship between the
$R$-matrix parameters describing a level and its observed width,
as discussed by \textcite[Sec.~XII.3, pp.~327-328]{Lan58}.
The sign is also important for establishing the uniqueness of the
alternative $R$-matrix parametrization given by \textcite{Bru02}.
We will elaborate on these points further in the Conclusions,
Sec.~\ref{sec:conclusions}.
While this sign appears to be positive in practice, a general proof for
positive energies is lacking and several authors have commented on
this point~\cite{Lan58,Bru02,Des10}.
Lane and Thomas did show that it is positive for negative energies
\cite[Eq.~(A.29), p.~351]{Lan58}, for positive energies in the JWKB
approximation \cite[Eq.~(A.19), p.~350]{Lan58}, and they also gave a
heuristic argument that it should be positive below the
Coulomb and/or angular momentum barriers \cite[Eq.~(A.32), p.~352]{Lan58}.
It is also straightforward to show that this sign is positive in the limits
of zero radius, infinite radius, zero energy, and infinite energy
(see Appendices~\ref{app:coulomb_rho} and~\ref{app:coulomb_e} below).

We have succeeding in proving that the energy derivative of the shift factor
is always positive for the repulsive Coulomb case.
We report results for the sign of the derivatives of the shift and
penetration factors, as well as the related amplitude and phase,
with respect to energy, angular momentum, and charge.
The results are obtained using the phase-amplitude parametrization of the
Coulomb functions and a little-known Nicholson-type integral representation
for the sum of squares of the regular and irregular Coulomb functions.
We also find that almost none of the results are generally valid
in the attractive Coulomb case.

This paper is organized as follows. We first review the relevant properties
of the Coulomb functions and then derive derive the monotonicity
results, with discussion and conclusions following. Appendices include
information on the theory of monotonic functions, further properties of
Coulomb functions, and additional integral relations for the
energy derivative of the shift factor.

\section{Overview of Coulomb functions}

\subsection{Definitions}

In terms of physical parameters, a Coulomb function $u$ satisfies
\begin{equation}
-\frac{\hbar^2}{2\mu}\frac{d^2u}{dr^2} +\frac{Z_1Z_2q^2}{r}u
+\frac{\hbar^2}{2\mu}\frac{\ell(\ell+1)}{r^2}u=Eu,
\label{eq:coul_r}
\end{equation}
where $r\ge 0$ is the radial coordinate, $E$ is the energy,
$\hbar^2/\mu$ is a positive constant,
$Z_1Z_2q^2/r$ is the Coulomb potential, and
${\hbar^2\ell(\ell+1)/(2\mu r^2)}$ is an effective potential corresponding
to the centrifugal or angular momentum barrier.
The quantity $\ell$ is the angular momentum quantum number and is a
non-negative integer in physical applications,
but unless otherwise indicated we will consider it to be a non-negative
continuous real parameter.
The quantity $Z_1Z_2q^2$ is the constant charge factor that is positive for a
repulsive Coulomb field, zero in the neutral case, and negative otherwise.
We will also assume $E>0$, unless otherwise indicated.
In terms of the dimensionless parameters $\rho$ and $\eta$,
we have $u(\ell,\eta,\rho)$ and this equation becomes
\begin{equation}
u''+\left[ 1-\frac{2\eta}{\rho}-\frac{\ell(\ell+1)}{\rho^2}\right]u=0,
\label{eq:coulomb_wave}
\end{equation}
where $\rho=kr$, $k=\sqrt{2\mu E/\hbar^2}$, $\eta k = Z_1Z_2q^2\mu/\hbar^2$,
and $'\equiv d/d\rho$.
The outgoing $u= H_\ell^+$ and incoming $u= H_\ell^-$ solutions
are given respectively by
\begin{subequations}
\begin{eqnarray}
 H_\ell^+ &=& G_\ell + iF_\ell \quad\quad {\rm and} \\
 H_\ell^- &=& G_\ell - iF_\ell,
\end{eqnarray}
\end{subequations}
where $G_\ell(\eta,\rho)\equiv G_\ell$ and $F_\ell(\eta,\rho)\equiv F_\ell$
are the irregular and regular Coulomb functions, respectively.

The logarithmic derivative of the outgoing solution is given by
\begin{equation}
L_\ell \equiv \frac{r}{H_\ell^+}\frac{dH_\ell^+}{dr} =
\rho\frac{H_\ell^{+\prime}}{H_\ell^+},
\end{equation}
with the real and imaginary parts defined to be
\begin{equation} \label{eq:l_s_p}
L_\ell \equiv S_\ell + i P_\ell,
\end{equation}
where $S_\ell$ and $P_\ell$ are the shift and penetration factors, respectively.
Note that for for $E\le 0$ we have $P_\ell = 0$.
It is also customary to define the asymptotic phase
\begin{equation}
\theta_\ell=\rho-\eta\log(2\rho)-\frac{1}{2}\ell\pi+\sigma_\ell,
\end{equation}
where $\sigma_\ell$ is the Coulomb phase shift defined in
Appendix~\ref{app:cphase_h}.
We also define the energy derivative $\partial E$
which is understood to be taken at fixed radius (i.e., with the
product $\eta\rho$ fixed):
\begin{subequations} \label{eq:partial_e}
\begin{eqnarray}
\frac{\partial}{\partial E} &=& \frac{2\mu r^2}{\hbar^2}\left(
  \frac{1}{2\rho}\frac{\partial}{\partial\rho}
  -\frac{\eta}{2\rho^2}\frac{\partial}{\partial\eta} \right) \\
  &=&\frac{\rho}{2E}\left( \frac{\partial}{\partial\rho}
  -\frac{\eta}{\rho}\frac{\partial}{\partial\eta} \right).
\end{eqnarray}
\end{subequations}

\subsection{Amplitude and phase}
\label{subsec:amp_phase}

It is possible to parametrize the Coulomb functions in terms of
an amplitude (or modulus) $A_\ell$ and phase
$\phi_\ell$~\cite{Whe37,Tha57,Lan58,Pro58}:
\begin{eqnarray}
A_\ell &=& (F_\ell^2 + G_\ell^2)^{1/2}, \\
\phi_\ell &=& \tan^{-1} F_\ell/G_\ell, \\
H_\ell^\pm &=& A_\ell\exp(\pm i\phi_\ell), \\
P_\ell &=& \frac{\rho}{A_\ell^2}, \quad {\rm and} \label{eq:p_a} \\
S_\ell &=& \frac{\rho A_\ell'}{A_\ell} = \frac{\rho(A_\ell^2)'}{2A_\ell^2},
  \label{eq:s_a}
\end{eqnarray}
where the Wronskian relation
\begin{equation}
H_\ell^+ H_\ell^{-\prime} - H_\ell^{+\prime} H_\ell^- =-2i
\end{equation}
has been used to derive Eq.~(\ref{eq:p_a}) from Eq.~(\ref{eq:l_s_p}).
The amplitude and phase obey the following differential equations:
\begin{eqnarray}
&&A_\ell^{\prime\prime} + \left[ 1-\frac{2\eta}{\rho}
  -\frac{\ell(\ell+1)}{\rho^2} \right]A_\ell
  -A_\ell^{-3} = 0 \quad {\rm and} \label{eq:A_difeq} \\
&&\phi_\ell' = A_\ell^{-2}. \label{eq:phi_prime}
\end{eqnarray}
In this work we make extensive use of square of $A_\ell$ and we will
refer to $A_\ell^2$ as ``the amplitude.''
A differential equation satisfied by $A_\ell^2$ is discussed below
in Subsec.~\ref{subsec:amplitude}.

R.G. Thomas derived an integral representation for $A_\ell^2$ that is useful
for establishing its monotonic properties
\cite[p.~350]{[{}][{.
There is a misprint in the third formula from the bottom of
p.~350, column~1 (un-numbered): an equals sign should be placed between
$(F^2+G^2)/(2\rho)$ and the $\int$ symbol.}]Lan58}, \cite{Pro58}:
\begin{subequations} \label{eq:a2}
\begin{align}
A_\ell^2 &= 2\rho\int_0^\infty dz \, e^{-2\rho z} \, Q(z), \quad {\rm where}
  \label{eq:a_qz} \\
Q(z) &= \exp(2\eta\tan^{-1}z)(1+z^2)^\ell \label{eq:q_2f1} \\
  & \times {}_2F_1(-\ell-i\eta,-\ell+i\eta,1;\frac{z^2}{1+z^2}) \nonumber \\
&= \exp(2\eta\tan^{-1}z)(1+z^2)^{i\eta} \label{eq:q_2} \\
  & \times {}_2F_1(\ell+1+i\eta,-\ell+i\eta,1;-z^2). \nonumber
\end{align}
\end{subequations}
The equivalence of the two expressions for $Q(z)$ results from
Pfaff and Euler transformations of the hypergeometric function.
This formula is also given in \textcite[Eq.~(12.5), p.~440]{[{}][{.
One of the factors of $-i\eta$ in the hypergeometric function in their
Eq.~(12.5) on p.~440 must be reversed in sign in order to agree with
Eq.~(\protect\ref{eq:q_2f1}) of the present work. Also, the $n=5$ line of their
Table~1 on p.~440 must contain an error because the asymptotic expansion
of $A_\ell$ only consists of even terms when $\eta=0$.}]Hul59},
but one of the factors of $-i\eta$ in their hypergeometric funtion must be
reversed in sign in order to agree with Eq.~(\ref{eq:q_2f1}).

Expressions such as Eq.~(\ref{eq:a2}) are known as
{\em Nicholson-type integrals}; further discussion is provided
below in Subsec.~\ref{subsec:amplitude}.
This equation appears to have been overlooked for over
half a century, but it is very useful in the present context.
The formula is based on a result given by \textcite{Erd38} that
expresses the product of two Whittaker functions as a Laplace transform,
which is applicable since we also have
\begin{equation}
\begin{split}
A_\ell^2 &= H_\ell^+ H_\ell^- \\
  &= e^{\pi\eta} \, W_{-i\eta,\ell+1/2}(-2i\rho) \, W_{i\eta,\ell+1/2}(2i\rho) ,
\end{split}
\end{equation}
where $W$ is the Whittaker function.

The particular hypergeometric function in Eq.~(\ref{eq:q_2f1}) may be
defined via
\begin{subequations} \label{eq:ft}
\begin{align}
& t = \frac{z^2}{1+z^2}, \\
& {}_2F_1(-\ell-i\eta,-\ell+i\eta,1;t) \equiv F(t) = \sum_{n=0}^\infty d_n t^n, \\
& d_0 = 1, \quad {\rm and} \\
& d_{n+1} = d_n \frac{\eta^2 +(n-\ell)^2}{(n+1)^2},
\end{align}
\end{subequations}
which is absolutely convergent for $|t|\le 1$. We also note that $F(t)$ is
real, positive, and monotonically increasing between $F(0)=1$ and
\begin{equation}
F(1) = \frac{2^{2\ell}e^{-\pi\eta}}{C_\ell^2(\eta) (2\ell+1)^2\Gamma(2\ell+1)},
\end{equation}
where $C_\ell(\eta)$ is defined in Appendix~\ref{app:coulomb_rho}.
The function $Q(z)$ is likewise positive for $0\le z <\infty$.
The integral representation of Erd{\' e}lyi
and the positivity of the integrand in certain circumstances
are also remarked upon by \textcite[p. 89, Eq.~(10a)]{Buc69}.

Equation~(\ref{eq:a_qz}) may be integrated by parts to
yield~\cite{Pro58,Lan58}:
\begin{equation}
A_\ell^2 = 1 + \int_0^\infty dz \, e^{-2\rho z} \, \frac{dQ}{dz}.
\label{eq:a_dqz}
\end{equation}
Further integrations by parts would yield an asymptotic expansion for
$A_\ell^2$ in terms of inverse powers of $\rho$.
We also have
\begin{subequations} \label{eq:dqdz}
\begin{eqnarray}
\frac{1}{Q}\frac{dQ}{dz} &=& \frac{2(\eta+\ell z)}{1+z^2} +
\frac{1}{F(t)} \frac{dF(t)}{dt}\frac{dt}{dz}, \\
\frac{dF(t)}{dt} &=& \sum_{n=1}^\infty nd_nt^{n-1}, \quad {\rm and} \\
\frac{dt}{dz} &=& \frac{2z}{(1+z^2)^2}.
\end{eqnarray}
\end{subequations}
Considering only the repulsive Coulomb case ($\eta>0$), we clearly have
$dQ/dz>0$, and hence $A_\ell^2>1$. By differentiating
Eq.~(\ref{eq:a_dqz}) with respect to $\rho$,
one can see that $(A_\ell^2)'<0$ and consequently
$S_\ell<0$~\cite{Pro58,Lan58}.
Further differentiation shows that all derivatives of $A_\ell^2$
have well-defined sign:
\begin{equation}
0 <  (-1)^n \left(\frac{d}{d\rho}\right)^n A_\ell^2 < \infty \quad\quad
  n=1,2,3,\ldots
\label{eq:a2n}
\end{equation}
This result shows that $A_\ell^2$ is a {\em Completely Monotonic} (CM) function
of $\rho$. Many of the conclusions reached in this paper follow
from this fact and are proven rather easily using the machinery
of CM functions. Some properties of CM functions are discussed in
Subsec.~\ref{app:monotonic:cm} of the Appendix;
additional details are available in the review article of \textcite{Mil01}.

Using Eqs.~(\ref{eq:a_qz}) and~(\ref{eq:q_2f1}),
\textcite{[{}][{.
Note that this reference defines the shift factor $S$ with the
opposite sign compared to modern conventions.
This paper also contains some errors.
In Sec.~III(c), Eq.~(25) is incorrect [compare to
Eq.~(\protect\ref{eq:shift_zero_e}) of the present work], as is the
statement that $S$ cannot be monotonic in energy.}]Pro58}
showed that $\partial (A_\ell^2)/\partial\eta > 0$; noting that
\begin{equation}
\frac{\partial P_\ell}{\partial E} = \frac{\rho}{2E} \left[
  \frac{A_\ell^2 - \rho (A_\ell^2)' +
  \eta\frac{\partial A_\ell^2}{\partial\eta} }{A_\ell^4} \right],
\end{equation}
it is clear that ${\partial P_\ell}/{\partial E} >0$.
These authors went on to show that ${\partial S_\ell}/{\partial \eta} <0$.
However, it does not appear to be feasible to extend their approach
to determine the sign of ${\partial S_\ell}/{\partial E}$.

Some additional properties of the Coulomb functions are discussed
in Appendices~\ref{app:cphase_h}-\ref{app:coulomb_e}.
It should be noted that $S_\ell$ is {\em not} monotonic in $\rho$:
from the formulas given in Appendix~\ref{app:coulomb_rho} it is clear that
$S_\ell'$ is negative for $\rho\rightarrow 0$ and positive for
$\rho\rightarrow\infty$.

\section{Energy derivative of \textit{L}}
\label{sec:dlde}

Using the differential equation with two different solutions
$O_1$ and $O_2$ with outgoing wave boundary conditions
(i.e., $O \propto H^+$) corresponding to energies $E_1$ and $E_2$ in
Eq.~(\ref{eq:coul_r}), one can show that
\begin{equation}
-\frac{\hbar^2}{2\mu}\frac{d}{dr}\left[\frac{O_1O_2}{r}(L_2-L_1)\right]=
  (E_2-E_1) O_1O_2.
\end{equation}
Note that the $\ell$ (angular momentum label) subscripts will be suppressed
from this point forward in this paper.
Upon integrating from $r=a$ to~$b$ with $a<b$ this becomes
\begin{equation}
-\frac{\hbar^2}{2\mu}\left[\frac{O_1O_2}{r}(L_2-L_1)\right]_a^b =
  (E_2-E_1)\int_a^b O_1O_2 \, dr.
\label{eq:int_o1o2}
\end{equation}
In the limit that $O_2\rightarrow O_1$, this becomes
\begin{equation}
-\frac{\hbar^2}{2\mu}\left[ \frac{O^2}{r}\frac{\partial L}{\partial E}
  \right]_a^b =\int_a^b O^2 \, dr,
\label{eq:dlde_ab}
\end{equation}
where $\partial E$ is taken at fixed radius as discussed above.
For bound states ($E<0$), $O$ is proportional to the exponentially-decaying
Whittaker function and one can take $b\rightarrow\infty$ with the surface
term at $r=b$ in the left-hand side of Eq.~(\ref{eq:dlde_ab}) vanishing;
see also \textcite[Eq.~(A.29), p.~351]{Lan58}:
\begin{equation}
\frac{\hbar^2}{2\mu}\left[\frac{O^2}{r}\frac{\partial L}{\partial E}\right]_a
  = \int_a^\infty O^2 \, dr.
\label{eq:dlde_a}
\end{equation}
Since $O(r)/O(a)$ is real, it follows that $\partial S/\partial E$
is positive for $E<0$~\cite{Lan58}.

It is not immediately obvious how to extend this result to positive
energies because $O(r)/O(a)$ is non-zero and oscillating for
large $r$ and it is also necessarily a complex quantity.
We attempted to find an integral expression with a positive-definite
integrand, analogous to Eq.~(\ref{eq:dlde_a}).
These efforts were not successful; some of the results found are
given in Appendix~\ref{app:integrals}.
We show here a successful approach to proving $\partial S/\partial E>0$
for the repulsive Coulomb case, using an integral expression with an integrand
that oscillates in sign with properties that allow a definitive sign
for the integral to be deduced.

Adopting $O=H^+=A\exp(i\phi)$ and changing the integration variable from
$r$ to $\rho$, we can write Eq.~(\ref{eq:dlde_ab}) in terms
of the amplitude and phase
\begin{equation}
-E\left[ e^{2i\phi} \frac{A^2}{\rho} \frac{\partial L}{\partial E}
  \right]^{\rho_b}_{\rho_a} =
  \int_{\rho_a}^{\rho_b} A^2 \, e^{2i\phi} \, d\rho .
\end{equation}
We next change the integration variable to $\psi$, noting that $\psi$ is
a monotonically-increasing function of $\rho$:
\begin{eqnarray}
&& \psi \equiv 2[\phi(\rho)-\phi(\rho_a)] , \label{eq:def_psi} \\
&& \psi' = 2A^{-2} , \label{eq:psi_prime} \\
&& \psi_b =  2[\phi(\rho_b)-\phi(\rho_a)] , \quad {\rm and} \\
&& -E\left[ e^{i\psi} \frac{A^2}{\rho}
  \frac{\partial L}{\partial E}\right]^{\psi_b}_0
  =  \frac{1}{2} \int_0^{\psi_b} A^4 e^{i\psi} \, d\psi ,
\label{eq:int_a4psi}
\end{eqnarray}
where Eq.~(\ref{eq:psi_prime}) follows from Eq.~(\ref{eq:phi_prime}). Using
\begin{equation}
\frac{d(A^4)}{d\psi} = \frac{2 A^2 (A^2)' }{\psi'} = A^4 (A^2)',
  \label{eq:dA4_dpsi}
\end{equation}
we can integrate by parts to find
\begin{equation}
\begin{split}
\left[ e^{i\psi} \left( -E\frac{A^2}{\rho} \right.\right.&\left.\left.
  \frac{\partial L}{\partial E}
  +\frac{i}{2}A^4 \right)\right]^{\psi_b}_0 \\
  &= \frac{i}{2}\int_0^{\psi_b} A^4 (A^2)' e^{i\psi} \, d\psi .
\label{eq:l_int1}
\end{split}
\end{equation}
Considering the large-$\rho$ behavior of the Coulomb quantities given
in Tables~\ref{tab:coulomb} and~\ref{tab:asymp_de} of
Appendix~\ref{app:coulomb_rho}, one can now take
$\psi_b\rightarrow\infty$ as $A^4(A^2)'\sim -{\eta}/{\rho^2}$ and
the integral is absolutely convergent:
\begin{equation}
\left[E\frac{A^2}{\rho} \frac{\partial L}{\partial E}
  -\frac{i}{2}A^4 \right]_{\rho_a} =
  \frac{i}{2}\int_0^{\infty} A^4 (A^2)' e^{i\psi} \, d\psi .
\label{eq:l_int2}
\end{equation}
Taking the real part of this expression yields
\begin{equation}
E\left[ \frac{A^2}{\rho} \frac{\partial S}{\partial E}\right]_{\rho_a} =
  -\frac{1}{2}\int_0^{\infty}  A^4 (A^2)' \sin(\psi) \, d\psi .
\label{eq:s_int_psi}
\end{equation}

Our strategy will be to use the fact that $A^2$ is a CM function
of $\rho$ (see Subsecs.~\ref{subsec:amp_phase} and~\ref{app:monotonic:cm})
to prove that certain integrals, such as the one appearing in
Eq.~(\ref{eq:s_int_psi}), have definite sign.
Since $A^4$ is the product of two CM functions (i.e., $A^2\times A^2$)
it is also CM. Furthermore, we have
\begin{equation}
\begin{split}
\left(-\frac{d}{d\psi}\right)^n A^4 =&
  \left(-\frac{A^2}{2}\frac{d}{d\rho}\right)^n A^4 >0 \\
& n=1,2,3,\ldots
\end{split}
\end{equation}
and thus $A^4$ is also a CM when considered as a function of $\psi$.
The quantity
\begin{equation}
-\frac{d A^4}{d\psi}=-A^4 (A^2)'
\end{equation}
appearing in the r.h.s of Eq.~(\ref{eq:s_int_psi}), which is the negative
of Eq.~(\ref{eq:dA4_dpsi}), is thus a CM function of $\psi$.
In particular, the fact that this quantity is positive and
monotonically decreasing allows one to conclude that the r.h.s.
of Eq.~(\ref{eq:s_int_psi}) is positive using reasoning given
in Appendix~\ref{app:monotonic}.
To summarize, the definite integral from zero to infinity of a CM function
multiplied by the sine or cosine function is positive,
provided the integral converges. We thus finally have
\begin{equation}
E\left[ \frac{A^2}{\rho} \frac{\partial S}{\partial E}\right]_{\rho_a} =
  -\frac{1}{2}\int_0^{\infty}  A^4 (A^2)' \sin(\psi) \, d\psi >0 ,
\end{equation}
and we can conclude that $\partial S/\partial E$ is indeed always positive
for $E>0$ and a repulsive Coulomb field.

This method can also provide information about ${\partial P}/{\partial E}$.
Starting from Eq.~(\ref{eq:int_a4psi}) and choosing $b$ such that
\begin{equation}
\psi_b=\psi_n=2\pi n \quad\quad n=1,2,3,...
\end{equation}
the range of integration becomes an integer multiple of the
period of $e^{i\psi}$ and the surface terms are simplified since
$e^{i\psi_n}=1$:
\begin{equation}
-E\left[ \frac{A^2}{\rho} \frac{\partial L}{\partial E}\right]^{\psi_n}_0
 =  \frac{1}{2}\int_0^{\psi_n} A^4 e^{i\psi} \, d\psi .
\end{equation}
Taking imaginary part, we have
\begin{equation}
\begin{split}
E\left[ \frac{A^2}{\rho} \frac{\partial P}{\partial E}\right]_{\rho_a} =&
  \,\,\frac{1}{2}\int_0^{\psi_n}  A^4 \sin(\psi) \, d\psi \\
  &+ E\left[ \frac{A^2}{\rho} \frac{\partial P}{\partial E}\right]_{\rho_n},
\end{split}
\end{equation}
where $\rho_n=\rho_b$ when $\psi_b=\psi_n$.
We cannot take $n\rightarrow\infty$ in this case since
$A^4\sim 1$ for large $\rho$ (at least without employing a regularization
procedure), but it is sufficient to consider $n$ to be very large such
that the asymptotic expansions of the Coulomb functions are applicable
(see Tables~\ref{tab:coulomb} and~\ref{tab:asymp_de} in
Appendix~\ref{app:coulomb_rho}):
\begin{equation}
\begin{split}
E\left[ \frac{A^2}{\rho} \frac{\partial P}{\partial E}\right]_{\rho_a} =&
  \,\,\frac{1}{2}\int_0^{\psi_n}  A^4 \sin(\psi) \, d\psi \\
  &+ \frac{1}{2}\left( 1 + \frac{2\eta}{\rho_n} + \ldots \right).
\label{eq:dpde1}
\end{split}
\end{equation}
Since $A^4$ is a CM function of $\psi$, we observe that both terms on the
r.h.s. of Eq.~(\ref{eq:dpde1})
are positive and we can conclude that $\partial P/\partial E >0$
(which has been derived previously using a different method~\cite{Pro58}).
In fact, we can do better because the surface term is non-zero
as $\rho_n\rightarrow\infty$:
\begin{equation}
\frac{\partial P}{\partial E} > \frac{\rho}{2EA^2}.
\end{equation}

It is also interesting to consider further integrations by parts. Since
\begin{equation}
\begin{split}
\int e^{\alpha x} f \, dx &= \sum_{k=0}^m (-1)^k\frac{e^{\alpha x}}{\alpha^{k+1}}
  \frac{d^k f}{dx^k} \\ &+ (-1)^{m+1} \int \frac{e^{\alpha x}}{\alpha^{m+1}}
  \frac{d^{m+1} f}{dx^{m+1}} \, dx
\end{split}
\end{equation}
for $m=0,1,2,\ldots$, Eq.~(\ref{eq:l_int2}) generalizes to
\begin{equation}
\begin{split}
\left[2E\frac{A^2}{\rho}\frac{\partial L}{\partial E}
  \right]_{\rho_a} &=
\left[ \sum_{k=0}^m i^{k+1}
  \left(\frac{d}{d\psi}\right)^k A^4 \right]_{\rho_a} \\
  & + \, i^{m+1} \int_0^{\infty} e^{i\psi}
  \left(\frac{d}{d\psi}\right)^{m+1} A^4 \, d\psi .
\end{split}
\end{equation}
Setting $m=1$ provides
\begin{equation}
\begin{split}
\left[2E\frac{A^2}{\rho}\frac{\partial L}{\partial E}
  \right]_{\rho_a} =& \left[ iA^4 - \frac{d(A^4)}{d\psi} \right]_{\rho_a} \\
  &-\int_0^\infty e^{i\psi} \frac{d^2(A^4)}{d\psi^2} \, d\psi
\label{eq:l_int3}
\end{split}
\end{equation}
and then taking the imaginary part gives
\begin{equation}
\left[2E\frac{A^2}{\rho}\frac{\partial P}{\partial E} -A^4
  \right]_{\rho_a} = -\int_0^\infty \sin(\psi)
  \frac{d^2(A^4)}{d\psi^2} \, d\psi .
\end{equation}
Since the r.h.s. of this equation must be negative, it provides
an upper-limit constraint on $\partial P/\partial E$:
\begin{equation}
\frac{\partial P}{\partial E} < \frac{\rho}{2E}A^2 ,
\label{eq:p_upper}
\end{equation}
This result could also be deduced from the imaginary part of
Eq.~(\ref{eq:l_int2}).
Taking the real part of Eq.~(\ref{eq:l_int3}) yields
\begin{equation}
\left[2E\frac{A^2}{\rho}\frac{\partial S}{\partial E} + \frac{d(A^4)}{d\psi}
  \right]_{\rho_a} = -\int_0^\infty \cos(\psi)
  \frac{d^2(A^4)}{d\psi^2} \, d\psi .
\end{equation}
Since the r.h.s. of this equation must be negative, this implies
\begin{equation}
\frac{\partial S}{\partial E} < -\frac{\rho}{2EA^2}
  \left[ \frac{d(A^4)}{d\psi} \right]_{\rho_a}
  = -\frac{\rho}{2E}A^2(A^2)'.
\end{equation}
The results of this section are summarized in the first two lines of
Table~\ref{tab:results_s_p}.

\begin{table}[tbh]
\caption{Summary of the result of Secs.~\protect\ref{sec:dlde}
and \protect\ref{sec:var_ell_charge} for the variation of $S$ and $P$
with $E$, $\ell$, and $\eta$. The second column is deduced from
the first using Eqs.~(\protect\ref{eq:p_a}) and~(\protect\ref{eq:s_a}).}
\label{tab:results_s_p}
\begin{ruledtabular}
\begin{tabular}{l@{\hspace{0.18in}}l}
${\scriptstyle 0} < \frac{\partial S}{\partial E} <
  -\frac{\rho A^2 (A^2)'}{2E}$ &
${\scriptstyle 0} < \frac{\partial S}{\partial E} <
  -\frac{\rho^2 S}{EP^2}$ \\[1ex]
$\frac{\rho}{2EA^2} < \frac{\partial P}{\partial E} < \frac{\rho A^2}{2E}$ &
$\frac{P}{2E} <  \frac{\partial P}{\partial E} < \frac{\rho^2}{2EP}$ \\[1ex]
${\scriptstyle (2\ell+1)}\frac{A^4}{\rho^2}
  \left[\frac{\rho(A^2)'}{2A^2}-\frac{1}{2}\right]
  < \frac{\partial S}{\partial \ell} < {\scriptstyle 0}$ &
$\frac{2\ell+1}{P^2}\left({\scriptstyle S}-\frac{1}{2}\right) <
  \frac{\partial S}{\partial \ell} < {\scriptstyle 0}$ \\[1ex]
$-\frac{(2\ell+1)A^2}{2\rho} < \frac{\partial P}{\partial\ell} <
  {\scriptstyle 0}$ &
$-\frac{2\ell+1}{2P} < \frac{\partial P}{\partial\ell} <
  {\scriptstyle 0}$ \\[1ex]
$\frac{2A^4}{\rho}\left[\frac{\rho(A^2)'}{2A^2}-\frac{1}{4}\right] <
  \frac{\partial S}{\partial \eta} < {\scriptstyle 0}$ &
$\frac{2\rho}{P^2}\left({\scriptstyle S}-\frac{1}{4}\right) <
  \frac{\partial S}{\partial \eta} < {\scriptstyle 0}$ \\[1ex]
$-{\scriptstyle A^2} < \frac{\partial P}{\partial\eta} < {\scriptstyle 0}$ &
$-\frac{\rho}{P} < \frac{\partial P}{\partial\eta} < {\scriptstyle 0}$
\end{tabular}
\end{ruledtabular}
\end{table}

\section{Variation of \textit{L} with angular momentum and charge}
\label{sec:var_ell_charge}

It is also interesting and feasible with the above approach to investigate
the variation of $L$ with the angular momentum $\ell$ and charge.
On page~414 of their article, \textcite{Pro58} stated that
$\partial (A^2)/\partial\ell>0$ (and consequently also
$\partial P/\partial\ell<0$) based on Eqs.~(\ref{eq:a_qz})
and ~(\ref{eq:q_2f1}) of the present paper,
but it is not clear how they arrived at that conclusion.
The other statements made by the authors in that paragraph follow simply
from the properties of $Q(z)$, but this one does not.
Assuming that $Q(z)$ is defined by Eq.~(\ref{eq:q_2f1}),
$\partial (A^2)/\partial\ell>0$ is true provided that
$\partial F(t)/\partial\ell>0$,
where $F(t)$ is the hypergeometric function defined by Eq.~(\ref{eq:ft}).
However, it is not always the case that $\partial F(t)/\partial\ell>0$.
Ref.~\cite{Pro58} was unable to find a result for $\partial S/\partial\ell$.

Using Eq.~(\ref{eq:coulomb_wave}) with two different solutions
$O_1$ and $O_2$ corresponding to angular momenta $l_1$ and $l_2$ but
with the same energy, one finds
\begin{equation}
\frac{d}{d\rho}\left[\frac{O_1O_2}{\rho}(L_2-L_1)\right]=
  (\ell_1+\ell_2+1)(\ell_2-\ell_1) \frac{O_1O_2}{\rho^2}.
\end{equation}
Integrating and taking $O_2\rightarrow O_1$ (considering $\ell$ to be a
continuous parameter) leads to
\begin{equation} \label{eq:dldl}
\left[\frac{O^2}{\rho}\frac{\partial L}{\partial\ell}\right]_{\rho_a}^{\rho_b} =
(2\ell+1)\int_{\rho_a}^{\rho_b} \frac{O^2}{\rho^2} \, d\rho.
\end{equation}
One can now take $\rho_b\rightarrow\infty$ and proceed as before:
\begin{equation} \label{eq:dldl_1}
\frac{2}{2\ell+1}
\left[\frac{A^2}{\rho}\frac{\partial L}{\partial\ell}\right]_{\rho_a} =
-\int_0^\infty \frac{A^4 e^{i\psi}}{\rho^2} \, d\psi .
\end{equation}
Noting that $\rho^{-2}$ is a CM function of $\rho$, and that hence
$A^4/\rho^2$ is likewise CM, we have
\begin{equation}
\begin{split}
\left(-\frac{d}{d\psi}\right)^n\frac{A^4}{\rho^2} =&
  \left(-\frac{A^2}{2}\frac{d}{d\rho}\right)^n\frac{A^4}{\rho^2} > 0 \\
& n=1,2,3,\ldots,
\end{split}
\end{equation}
and we can conclude immediately that
\begin{equation}
\frac{\partial S}{\partial \ell}<0 \quad {\rm and} \quad
  \frac{\partial P}{\partial\ell}<0,
\end{equation}
using the methods of Appendix~\ref{app:monotonic}.
Integrating Eq.~(\ref{eq:dldl_1}) by parts twice yields
\begin{equation}
\begin{split}
\frac{2}{2\ell+1}
\left[\frac{A^2}{\rho}\frac{\partial L}{\partial\ell}\right]_{\rho_a} =&
\left[-i\frac{A^4}{\rho^2}+\frac{A^6}{\rho^3}
\left(\frac{\rho(A^2)'}{A^2}-1\right) \right]_{\rho_a} \\ &+
\int_0^\infty \frac{d^2}{d\psi^2}\left(\frac{A^4}{\rho^2}\right)e^{i\psi}
\, d\rho ,
\end{split}
\end{equation}
which shows
\begin{equation}
\frac{\partial S}{\partial \ell}>(2\ell+1)\frac{A^4}{\rho^2}\left[
\frac{\rho(A^2)'}{2A^2}-\frac{1}{2}\right]
\end{equation}
and
\begin{equation}
 \frac{\partial P}{\partial\ell} > -\frac{(2\ell+1)A^2}{2\rho}.
\end{equation}

The variation of $L$ with charge can be studied using this procedure via
the Coulomb parameter $\eta$. This results in
\begin{equation}
\frac{d}{d\rho}\left[\frac{O_1O_2}{\rho}(L_2-L_1)\right]=
  2(\eta_2-\eta_1) \frac{O_1O_2}{\rho}.
\end{equation}
Upon integrating and taking $O_2\rightarrow O_1$,
\begin{equation} \label{eq:dldeta}
\left[\frac{O^2}{\rho}\frac{\partial L}{\partial\eta}\right]_{\rho_a}^{\rho_b} =
  2\int_{\rho_a}^{\rho_b} \frac{O^2}{\rho} \, d\rho.
\end{equation}
Proceeding as above, we have
\begin{equation} \label{eq:dldeta_1}
\left[\frac{A^2}{\rho}\frac{\partial L}{\partial\eta}\right]_{\rho_a} =
  -\int_0^\infty \frac{A^4 e^{i\psi}}{\rho} \, d\psi .
\end{equation}
Noting that $\rho^{-1}$ is a CM function of $\rho$ and thus
\begin{equation}
\begin{split}
\left(-\frac{d}{d\psi}\right)^n\frac{A^4}{\rho} =&
  \left(-\frac{A^2}{2}\frac{d}{d\rho}\right)^n\frac{A^4}{\rho} > 0, \\
& n=1,2,3,\ldots,
\end{split}
\end{equation}
we conclude
\begin{equation}
\frac{\partial S}{\partial\eta} < 0 \quad {\rm and} \quad
\frac{\partial P}{\partial\eta} < 0,
\end{equation}
confirming the findings of Ref.~\cite{Pro58}.
Integrating Eq.~(\ref{eq:dldeta_1}) by parts twice yields
\begin{equation}
\begin{split}
\left[\frac{A^2}{\rho}\frac{\partial L}{\partial\eta}\right]_{\rho_a} =&
\left[-i\frac{A^4}{\rho}+\frac{A^6}{2\rho^2}
\left(\frac{2\rho(A^2)'}{A^2}-1\right) \right]_{\rho_a} \\ &+
\int_0^\infty \frac{d^2}{d\psi^2}\left(\frac{A^4}{\rho}\right)e^{i\psi}
\, d\rho ,
\end{split}
\end{equation}
which shows
\begin{equation}
\frac{\partial S}{\partial \eta}>\frac{2A^4}{\rho}\left[
\frac{\rho(A^2)'}{2A^2}-\frac{1}{4}\right]
\end{equation}
and
\begin{equation}
 \frac{\partial P}{\partial\eta} > -A^2.
\end{equation}
The results of this section are summarized in Table~\ref{tab:results_s_p}.

\section{Variation of the amplitude and phase with \textit{E},
  \texorpdfstring{$\bm{\ell}$}{l}, and \texorpdfstring{$\bm{\eta}$}{eta} }
\label{sec:var_amp_phase}

It is also expected that the amplitude and phase depend monotonically on
$E$, $\ell$, and $\eta$.
Noting $P=\rho/A^2$, the variations of squared amplitude $A^2$
with $\ell$ and $\eta$ are easily found to be
opposite of those already derived for $P$:
$\partial A^2/\partial\ell >0$ and $\partial A^2/\partial\eta>0$.
The latter result can also be shown by differentiating
Eq.~(\ref{eq:a_qz}) with $Q(z)$ given by Eq.~(\ref{eq:q_2f1})~\cite{Pro58}.
For the energy variation of the amplitude,
we have using Eq.~(\ref{eq:partial_e})
\begin{equation}
\frac{\partial A^2}{\partial E} = \frac{\rho}{2E}\left[ (A^2)'
  - \frac{\eta}{\rho}\frac{\partial A^2}{\partial \eta} \right] .
\end{equation}
Since $(A^2)'<0$ and $\partial A^2/\partial\eta>0$, we can also conclude that
$\partial A^2/\partial E<0$.

The variation of the phase $\phi$ with these parameters can also be related to
those for $P$. Noting that
\begin{equation}
\frac{d\phi}{dr}=\frac{P}{r} \quad {\rm and~hence} \quad
  [\phi]_a = \int_0^a \frac{P}{r} \, dr,
\end{equation}
we have
\begin{equation}
\left[\frac{\partial\phi}{\partial X}\right]_a = 
  \int_0^a \frac{\partial P}{\partial X}\frac{dr}{r},
\end{equation}
where $X = E$, $\ell$, or $\eta$, and the variation of $\phi$ with these
parameters is seen to be in the same direction as it is for $P$.
Note that in the case of $\ell=0$ for $\partial\phi/\partial\ell$
the integrand has a logarithmic singularity as $r\rightarrow 0$,
but the integral is still convergent.
The results of this section are summarized in Table~\ref{tab:results_a_phi}.

\begin{table}[tbh]
\caption{Summary of the result of Sec.~\protect\ref{sec:var_amp_phase}.}
\label{tab:results_a_phi}
\begin{ruledtabular}
\begin{tabular}{@{\hspace{0.25in}}l@{\hspace{0.5in}}l@{\hspace{0.5in}}%
  l@{\hspace{0.25in}}}
$\frac{\partial A^2}{\partial E} <0$ & $\frac{\partial A^2}{\partial\ell}>0$ &
  $\frac{\partial A^2}{\partial\eta}>0$ \\
$\frac{\partial\phi}{\partial E} >0$ & $\frac{\partial\phi}{\partial\ell}<0$ &
  $\frac{\partial\phi}{\partial\eta}<0$
\end{tabular}
\end{ruledtabular}
\end{table}

\section{Discussion}

We have limited our consideration to the repulsive Coulomb field
($\eta>0$) in this work. In this section we will briefly consider the
attractive Coulomb field and then in more detail the neutral case.
We next provide further discussion of the amplitude $A^2$,
followed by a brief review of negative energies.
In Fig.~\ref{fig:shift} we show shift factor $S(E)$ for the repulsive,
neutral, and attractive cases ($Z_1Z_2 = 1$, 0, and $-1$, respectively).
We have also assumed $\ell=0$, $q$ to be the fundamental charge,
$\mu$ to be the nucleon-nucleon reduced mass, and a radius of 2~fm.
The repulsive case shows the expected results:
$S<0$ and $\partial S/\partial E >0$ for all energies.

\subsection{The attractive Coulomb case}

\begin{figure}
\includegraphics[width=\columnwidth]{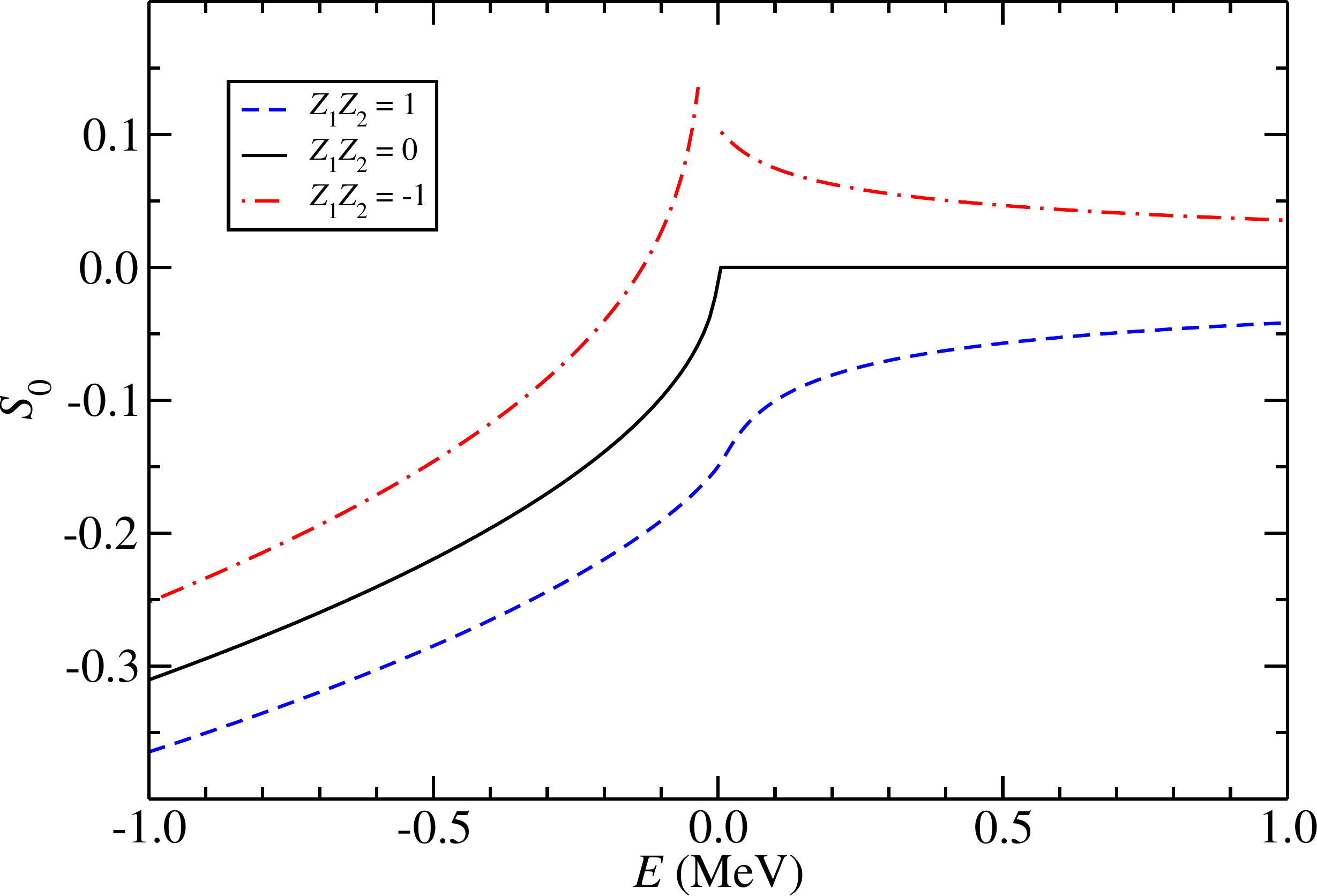}
\caption{(Color online) The $\ell=0$ shift factor versus energy for
the repulsive (blue dashed curve), uncharged (black solid curve),
and attractive (red dot-dashed curve) cases.
Additional details are provided in the text.} \label{fig:shift}
\end{figure}

In the case of an attractive Coulomb field,
the amplitude $A^2$ is no longer guaranteed to be a CM function of $\rho$
because according to Eq.~(\ref{eq:dqdz}) $dQ/dz$ is not necessarily positive.
Consequently, very few of the results from the repulsive case are generally
valid for $\eta<0$.
We can conclude that $A^2>0$ and hence $P>0$ from Eqs.~(\ref{eq:a_qz})
and~(\ref{eq:q_2f1}).
For the particular case with $\ell=0$ plotted in Fig.~\ref{fig:shift},
it can be seen that $\partial S/\partial E<0$ for $E>0$.

\subsection{The neutral case}

In the neutral or uncharged case, we have $\eta=0$ and the amplitude is
given by
\begin{equation} \label{eq:a2_neutral}
A^2 = \frac{\pi}{2}\rho\left[ J^2_{\ell+1/2}(\rho)+Y^2_{\ell+1/2}(\rho)\right],
\end{equation}
where $J$ and $Y$ are the regular and irregular Bessel functions.
It is convenient in this case to use form of $Q(z)$ given by
Eq.~(\ref{eq:q_2}), which becomes
\begin{equation} \label{eq:q_uncharged}
Q(z) = {}_2F_1(-\ell,\ell+1,1;-z^2).
\end{equation}
Following \textcite{Pro58}, Eq.~(\ref{eq:a_qz}) may then be integrated
termwise to yield
\begin{equation} \label{eq:a_3f0}
A^2 = {}_3F_0(-\ell,\ell+1,\frac{1}{2};-\rho^{-2}),
\end{equation}
where ${}_3F_0$ is a generalized hypergeometric function, and it
is assumed that both hypergeometric functions may be represented by
their canonical power series.
If $\ell$ is a non-negative integer (the case for physical problems),
then the hypergemeotric functions in Eqs.~(\ref{eq:q_uncharged})
and~(\ref{eq:a_3f0}) are represented by series that terminate and there
are no questions of convergence.
Otherwise, the ${}_2F_1$ in Eq.~(\ref{eq:q_uncharged}) cannot
be represented by its series when $z>1$ and the
series for ${}_3F_0$ in Eq.~(\ref{eq:a_3f0}) is a non-convergent
asymptotic expansion, equivalent to Eq.~13.75(1) of \textcite[p.~449]{Wat44}.

Alternatively, one may utilize the fact that Eq.~(\ref{eq:q_uncharged})
is a representation of the Legendre function $\tilde{P}_\ell$
(a polynomial if $\ell$ is a non-negative
integer)~\cite[Eq.~15.4.16, p.~562]{Abr65}
\begin{equation}
{}_2F_1(-\ell,\ell+1,1;-z^2) = \tilde{P}_\ell(1+2z^2),
\end{equation}
to write
\begin{equation}
A^2 = 2\rho \int_0^\infty dz \, e^{-2\rho z} \tilde{P}_\ell(1+2z^2).
\end{equation}
This is a known integral representation for Eq.~(\ref{eq:a2_neutral})
deduced by \textcite[Eq.~(11.6), p.~588]{Har73} using a different approach.

In the neutral case, the monotonicity results are essentially
unchanged from the repulsive charge case,
since $dQ/dz\ge 0$ [see Eq.~(\ref{eq:dqdz})].
Note that $dQ/dz=0$ only occurs when $\ell=0$, in which case we have
$Q=1$ and $A^2=1$, which leads to $P=\rho$, $S=0$, and $\phi=\rho$ for $E>0$.
This shift factor is plotted in Fig.~\ref{fig:shift}. 
In this case, the monotonicity properties are trivial may be deduced by
inspection. In particular, we note that $\partial S/\partial E=0$ for
$\ell=0$ and positive energy.

\subsection{Further discussion of the amplitude \texorpdfstring{$A^2$}{A2}}
\label{subsec:amplitude}

Our finding that $A^2 = F_\ell^2 + G_\ell^2$ is CM for the case of a
repulsive Coulomb field is a generalization
of the result that ${\rho[J_\nu^2(\rho)+Y_\nu^2(\rho)]}$ is CM~\cite{Lor63}.
The key to proving that $A^2$ is CM is the Laplace transform
representation given by Eq.~(\ref{eq:a2}).
Integral representations for the sum of squares of linearly-independent
solutions to an ordinary second-order differential equation,
such as Eq.~(\ref{eq:a2}), are known as {\em Nicholson-type integrals}
and may considered to be generalizations of ${\sin^2\rho+\cos^2\rho=1}$.
These representations are often useful for establishing monotonicity
properties of special functions~\cite{Lor63,Dur78,Mul86}, as has been the
case in the present work.

\textcite{Har61,Har73} has studied the differential equation
\begin{equation} \label{eq:hartman}
u'' +[c+s(\rho)]u=0,
\end{equation}
where $c$ is a positive constant and $s\rightarrow 0$ as
$\rho\rightarrow\infty$.
He has shown that there are always solutions $x$ and $y$ to
Eq.~(\ref{eq:hartman}) with unit Wronskian such that the generalized amplitude
$A^2={x^2+y^2}\rightarrow 1$ as $\rho\rightarrow\infty$ and,
if  $-s(\rho)$ is CM, the generalized amplitude $A^2$ is CM.
Since Eq.~(\ref{eq:coulomb_wave}) is of this form with both
the repulsive Coulomb potential and the centrifugal barrier
making CM contributions to $-s(\rho)$, this provides an alternate
proof that $A^2$ is CM for the repulsive Coulomb case.
It is also clear from this perspective that we are unable to draw
general conclusions regarding the monotonicity of
$A^2$ for an attractive Coulomb potential.

We finish the discussion of the amplitude by deriving its asymptotic
expansion for large $\rho$.
Leading asymptotic expansions for $A$ have been given by \textcite{Hul59}
that lack general formulas for the coefficients.
The asymptotic expansion for $A^2$ turns out to be considerably simpler.
If $u$ and $v$ are solutions of Eq.~(\ref{eq:coulomb_wave}),
their product $w=uv$ satisfies the Appell equation, a third-order homogeneous
linear differential equation~\cite[p.~560, Eq.~(2.23)]{Har73}, which
in our case reads
\begin{equation} \label{eq:appell}
\begin{split}
w'''+&4\left[1-\frac{2\eta}{\rho}-\frac{\ell(\ell+1)}{\rho^2}\right]w' \\
&+2\left[\frac{2\eta}{\rho^2}+\frac{2\ell(\ell+1)}{\rho^3}\right]w=0.
\end{split}
\end{equation}
Any linear combinations of such solutions, including $A^2=F_\ell^2+G_\ell^2$,
is likewise a solution of Eq.~(\ref{eq:appell}).
Assuming an expansion of the form
\begin{subequations} \label{eq:a_asymp}
\begin{equation}
A^2 \sim \sum_{k=0}^\infty \frac{a_k}{\rho^k}
\end{equation}
with $a_0=1$ and substituting into Eq.~(\ref{eq:appell}) leads to the
following result for the coefficients:
\begin{align}
&a_0 = 1, \\
&a_1 = \eta, \qquad \mbox{and for $k\ge 1$} \\
&a_{k+1} = \eta\frac{2k+1}{k+1}a_k \\ 
 &\quad\quad +\frac{k(2\ell+k+1)(2\ell-k+1)}{4(k+1)}a_{k-1} . \nonumber
\end{align}
\end{subequations}
Considering that any solution of Eq.~(\ref{eq:appell}) must be a linear
combination of $F_\ell^2$, $G_\ell^2$, and $F_\ell G_\ell$ and the leading
asymptotic expansions of these possibilities, it is clear that
Eq.~(\ref{eq:a_asymp}) is in fact the asymptotic expansion of $A^2$.
If $\eta=0$ (i.e., the neutral case), the expansion only contains even terms
and is equivalent to Eq.~(\ref{eq:a_3f0}).
If $\ell$ is also a non-negative integer (the case for physical problems),
the series terminates.
Equation (\ref{eq:a_asymp}) is the generalization to
the Coulomb case of the asymptotic series for ${J_\nu^2+Y_\nu^2}$ given by
Eq.~13.75(1) of \textcite[p.~449]{Wat44}.

\subsection{Negative energies}

In the case of negative energies, the Coulomb functions satisfy
\begin{equation}
u''+\left[ -1-\frac{2\eta}{\rho}-\frac{\ell(\ell+1)}{\rho^2}\right]u=0,
\end{equation}
where $\rho=kr$, $k=\sqrt{-2\mu E/\hbar^2}$, $\eta k = Z_1Z_2q^2\mu/\hbar^2$,
and $'\equiv d/d\rho$.
We will consider the solution given by the
exponentially-decaying Whittaker function $W_{-\eta,\ell+1/2}(2\rho)\equiv W$
and the shift factor that is given for negative energies by
\begin{equation}
S=\rho\frac{W'}{W}.
\end{equation}
Adapting Eqs.~(\ref{eq:dlde_a}), (\ref{eq:dldl}), and (\ref{eq:dldeta})
to negative energies, we have
\begin{eqnarray}
\left[\frac{W^2}{\rho}\frac{\partial S}{\partial E}\right]_{\rho_a} &=&
  -\frac{1}{E} \int_{\rho_a}^\infty W^2 \, d\rho , \\
\left[\frac{W^2}{\rho}\frac{\partial S}{\partial\ell}\right]_{\rho_a} &=&
  -(2\ell+1)  \int_{\rho_a}^\infty \frac{W^2}{\rho^2} \, d\rho , \quad {\rm and}\\
\left[\frac{W^2}{\rho}\frac{\partial S}{\partial\eta}\right]_{\rho_a} &=&
  -2  \int_{\rho_a}^\infty \frac{W^2}{\rho} \, d\rho .
\end{eqnarray}
These equations show $\partial S/\partial E>0$, $\partial S/\partial\ell<0$,
and $\partial S/\partial\eta<0$ for negative energies, regardless of whether
the Coulomb potential is repulsive, attractive, or zero.
These results have been noted previously --
see the discussion of Eq.~(\ref{eq:dlde_a}) above regarding 
$\partial S/\partial E$ and \textcite[Sec.~IV]{Pro58}.
Also, all of the shift factors plotted in Fig.~\ref{fig:shift} are
consistent with $\partial S/\partial E>0$ for $E<0$.
One should be aware that, for the attractive Coulomb case,
the shift factor has singularities for slightly negative energies due
to zeros of the Whittaker function.

In the absence of the Coulomb potential, the negative-energy solutions
are modified Bessel functions. \textcite{Gol58} showed that an
amplitude and phase parametrization can be implemented in this situation.
Here, the solutions depend exponentially on the ``phase,''
as opposed to the sinusoidal dependence used for positive energies.
Presumably, this description could be extended to include the Coulomb potential
and describe the negative-energy Whittaker function solutions.

\section{Conclusions}
\label{sec:conclusions}

We have studied the derivatives of the shift and
penetration factors, as well as the related amplitude and phase,
with respect to energy, angular momentum, and charge.
For the cases of neutral or repulsive Coulomb fields, we find definitive
results for the signs of these quantities, as summarized in
Tables~\ref{tab:results_s_p} and~\ref{tab:results_a_phi}.
In particularly, we have succeeded in proving $\partial S/\partial E>0$,
a result that has been long thought to be true,
but for which a general proof was lacking.

The fact that $\partial S/\partial E>0$ for positive energies and a
repulsive or neutral Coulomb field has implications for the $R$-matrix
description of nuclear reactions. When relating $R$-matrix reduced width
amplitudes to physical quantities, one is presented with the factor
\begin{equation} \label{eq:nfactor}
N^{-1} = 1+\sum_c \gamma^2_{\lambda c} \frac{\partial S_c}{\partial E},
\end{equation}
where $c$ is the channel label and $\gamma_{\lambda c}$ are the
reduced width amplitudes.
For an unbound state in the one-level approximation, the observed partial
width is given by \textcite[Eqs.~(3.5) and~(3.6), p. 327]{Lan58},
\begin{equation} \label{eq:Gamma}
\Gamma_{\lambda c} = 2NP_c\gamma^2_{\lambda c}.
\end{equation}
The Thomas approximation~\cite{Tho51} has been employed here, which
assumes that $S_c(E)$ may be replaced by its first-order
Taylor series.
Knowledge that $\partial S/\partial E>0$ ensures that $N>0$ and that
the observed partial width is non-negative, a requirement for
a physically-reasonable partial width.
In the case of a bound level, the factor $N$ defined by
Eq.~(\ref{eq:nfactor}) also arises.
In this situation, $N$ changes the normalization volume of the
wave function from inside the channel surfaces to all space
\cite[Sec.~IV.7, p.~280; Eqs.~(A.29) and~(A.30), p.~351]{Lan58}.
For this case, $N$ was already known to be positive.
The description of the physical properties of bound and unbound levels
may be unified by considering the complex poles of the
scattering matrix~\cite{Hal87}. In this approach, a similar normalization
factor containing $\partial L/\partial E$ naturally
appears in the residues of the scattering matrix poles.
If the level is narrow such that the pole is near the real energy axis, this
normalizaton factor becomes equivalent to Eq.~(\ref{eq:nfactor})
in the one-level approximation.
Although less fundamental than partial widths defined via the residues
of the poles of the scattering matrix, Eqs.~(\ref{eq:nfactor})
and~(\ref{eq:Gamma}) may serve as a practical definition of the
observed partial width in $R$-matrix theory.

\textcite{Bru02} has given an alternative parametrization of $R$-matrix
theory that utilizes level energies and reduced width amplitudes that are
more closely connected to the observed resonance energies and partial
widths than in the standard parametrization~\cite{Lan58}.
The present result that  $\partial S/\partial E>0$ is sufficient to prove that
the alternative parameters have a one-to-one relationship to the standard
parameters and ensures that the alternative parametrization is well defined
and fully equivalent to the standard parametrization.
Further information on this point is provided by Eq.~(45) of Ref.~\cite{Bru02}
and that equation's surrounding discussion.

The results given in this paper follow from the Nicholson-type integral
representation of the amplitude $A^2$ given by Eq.~(\ref{eq:a2}).
When the Coulomb field is repulsive or absent, we find that $A^2$ is a
CM function of $\rho$, which leads to definitive monotonicity properties
for the shift and penetration factors.
Considering the work of \textcite{Har61,Har73} on the theory of
differential equations, it is apparent that any central potential
that is a CM function of radius will give analogous results.
To be explicit, a CM potential is necessarily repulsive and
monotonically decreasing with radius, with the signs of higher
derivatives prescribed according to Eq.~(\ref{eq:CM}).
An attractive potential cannot be CM and almost none of the conclusions
of this paper apply in this case.

\begin{acknowledgments}
This work was carried out under the auspices of the U.S. Department of Energy,
under Grants No.~DE-FG02-88ER40387 and No.~DE-{NA0002905} at Ohio University
and Grant No.~DE-AC52-06NA25396 at Los Alamos National Laboratory.

\end{acknowledgments}

\appendix

\section{Some results concerning monotonic functions}
\label{app:monotonic}

\begin{table*}[tbh]
\caption{Limiting forms of various Coulomb quantities for small and
large $\rho$.
The complete asymptotic expansion of $A_\ell^2$ for large $\rho$ is given
in the text by Eq.~(\protect\ref{eq:a_asymp}).
A refined small-$\rho$ expansion for $S_\ell$ is given
in the text by Eq.~(\protect\ref{eq:s_small_rho}).}
\label{tab:coulomb}
\begin{tabular}{c@{\hspace{0.25in}}l@{\hspace{0.5in}}l}
\hline\hline
quantity & \multicolumn{1}{c}{$\rho\rightarrow 0$} &
  \multicolumn{1}{c}{$\rho\rightarrow\infty$} \\ \hline \rule{0ex}{4.0ex}
$H_\ell^+$ &
  $[\rho^\ell(2\ell+1)C_\ell(\eta)]^{-1}+\ldots
  +i\left[\rho^{\ell+1}C_\ell(\eta)+\ldots\right]$
  & $\exp(i\theta_\ell) \left[ 1 + \frac{\eta}{2\rho}
  + i\frac{\eta^2+\ell(\ell+1)}{2\rho} + \ldots \right]$ \\
$A_\ell^2$ &
  $[\rho^\ell(2\ell+1)C_\ell(\eta)]^{-2}$ +\ldots &
  $1 + \frac{\eta}{\rho} + \frac{3\eta^2+\ell(\ell+1)}{2\rho^2} + \ldots$ \\
$\phi_\ell$ &
  $\rho^{2\ell+1}(2\ell+1)C_\ell^2(\eta)$ +\ldots &
  $\theta_\ell + \frac{\eta^2+\ell(\ell+1)}{2\rho} + \ldots$ \\
$P_\ell$ &
  $\rho^{2\ell+1}[(2\ell+1)C_\ell(\eta)]^2$ +\ldots &
  $\rho - \eta - \frac{\eta^2+\ell(\ell+1)}{2\rho} + \ldots$ \\
$S_\ell$ & $-\ell$ +\ldots &
  $-\frac{\eta}{2\rho} -\frac{2\eta^2+\ell(\ell+1)}{2\rho^2} + \ldots$ \\
\hline\hline
\end{tabular}
\end{table*}

We summarize here some aspects of monotonic functions that are
of use in this paper.

\subsection{Integrals
  \texorpdfstring{of the type $\bm{\int_0^{2\pi m} f(x)\sin(x)\, dx}$}{} }
\label{app:monotonic:integrals}

Consider the integral
\begin{equation}
I = \int_0^{2\pi} f(x) \sin(x) \, dx ,
\end{equation}
where $f(x)>0$ and $f'(x)<0$ for $x>0$. The integral can be split and rewritten
as an integral from~0 to~$\pi$:
\begin{eqnarray}
I &=& \int_0^{\pi} f(x) \sin(x) \, dx + \int_{\pi}^{2\pi} f(x) \sin(x) \, dx \\
  &=& \int_0^{\pi} \left[ f(x)-f(x+\pi) \right] \sin(x) \, dx .
\end{eqnarray}
Since the conditions on $f(x)$ imply $f(x)-f(x+\pi) >0$ and $\sin(x) >0$ for
$0<x<\pi$ we can conclude that $I>0$. The same result holds if the
integration range is extended by an integer multiple $m$ of $2\pi$:
\begin{equation}
\int_0^{2\pi m}  f(x) \sin(x) \, dx >0 \quad m=1,2,3,\ldots ,
\end{equation}
including for $m\rightarrow\infty$, provided the integral converges.
Note that an analogous conclusion {\em cannot} in general be drawn for
$\int_0^{2\pi} f(x)\cos(x) \, dx$, but in may be possible to draw
conclusions using integration by parts -- depending on the sign of
$f''(x)$ (see Subsec.~\ref{app:monotonic:cm} below).

\subsection{Completely monotonic functions}
\label{app:monotonic:cm}

A function $f(x)$ is said to be {\em Completely Monotonic} (CM) if
\begin{equation} \label{eq:CM}
0 \le  (-1)^n \left(\frac{d}{dx}\right)^n f(x) < \infty \quad
\end{equation}
for all $x>0$ and $n=0,1,2,\ldots$
The properties of CM functions are reviewed in Ref.~\cite{Mil01}.
Besides the definition, the feature of CM functions that is particularly
useful for this work the fact that the product of two CM functions is
also a CM function.
A consequence of this property that we will utilize is that
\begin{equation}
h(x) = \left[ -g(x) \frac{d}{dx} \right]^n f(x) \quad\quad n=0,1,2,\ldots
\end{equation}
is a CM function if $f(x)$ and $g(x)$ are CM.
The definition Eq.~(\ref{eq:CM}) allows $f(x)$ to be a non-negative constant;
we exclude this case which eliminates the possibility that $(d/dx)^n f(x)$
in Eq.~(\ref{eq:CM}) can be zero \cite[Appendix~I, p.~71-72]{Lor63}.

We now consider a CM $f(x)$ and integrals over ${0<x<2\pi m}$ where
$m=1,2,3,\ldots$
Using the results of Subsec.~\ref{app:monotonic:integrals},
we can immediately conclude that
\begin{equation}
\int_0^{2\pi m}  f(x) \sin(x) \, dx >0 .
\end{equation}
We now also have an analogous result for the cosine integral:
\begin{align}
\int_0^{2\pi m} f(x)\cos(x) \, dx =& \left[ f(x)\sin(x) \right]_0^{2\pi m} \\
  &-\int_0^{2\pi m} \frac{df}{dx}\sin(x) \, dx \nonumber \\
=& -\int_0^{2\pi m} \frac{df}{dx}\sin(x) \, dx \\
>&0,
\end{align}
where the assumption that the original integral is convergent allows
one to conclude the surface term vanishes and we have used
the fact that $-df/dx$ is a CM function.
Similar reasoning has been used in Ref.~\cite{Tuc06} to derive
sufficient conditions for a Fourier sine or cosine transform to be
positive.

\section{Coulomb phase shift and \texorpdfstring{$\bm{h(\eta)}$}{h(eta)}}
\label{app:cphase_h}

The Coulomb phase shift $\sigma_\ell$ is defined by
\begin{equation} \label{eq:coul_phase}
e^{2i\sigma_\ell} = \frac{\Gamma(1+\ell+i\eta)}{\Gamma(1+\ell-i\eta)} =
  \frac{(\ell+i\eta)\ldots (1+i\eta)}{(\ell-i\eta)\ldots (1-i\eta)}
  e^{2i\sigma_0},
\end{equation}
with the derivative of $\sigma_\ell$ is given by
\begin{eqnarray}
\frac{d\sigma_\ell}{d\eta} &=& \frac{1}{2}\left[ \Psi(1+\ell+i\eta)
  + \Psi(1+\ell-i\eta) \right] , \\
\frac{d\sigma_{\ell >0}}{d\eta} &=& \frac{d\sigma_0}{d\eta}
  +\sum_{m=1}^{\ell} \frac{m}{m^2+\eta^2} , \label{eq:sig_ell} \\
\frac{d\sigma_0}{d\eta} &=& \frac{1}{2}\left[ \Psi(1+i\eta) + \Psi(1-i\eta)
  \right] , \quad {\rm and} \\
&\equiv& h(\eta) + \log(\eta), \label{eq:def_h_eta}
\end{eqnarray}
where $\Psi$ is the digamma function.
Note that the final part of Eq.~(\ref{eq:coul_phase}) and
Eq.~(\ref{eq:sig_ell}) are only applicable when $\ell$ is a
non-negative integer.
Equation~(\ref{eq:def_h_eta}) serves to define, for a repulsive Coulomb field,
the auxiliary function $h(\eta)$ that also arises in the series expansion
of the irregular Coulomb functions and
in effective range theory~\cite{Bet49,Jac50,Hae77}.
Note that $h(\eta)$ is real when $\eta$ is real and positive and that it
is also given by~\cite{Bet49,Jac50}
\begin{equation}
h(\eta) = -\log\eta -\gamma+\eta^2 \sum_{k=1}^{\infty}\frac{1}{k(k^2+\eta^2)},
\end{equation}
where $\gamma=0.57721566\ldots$ is Euler's constant.

It appears that $h(\eta)$ is a completely monotonic function of $\eta$,
but we have been unable to prove this.
Using Eq.~6.3.21 of \textcite[p.~259]{Abr65} and the properties of the digamma
function, we have found the following representation for $h(\eta)$:
\begin{subequations}
\begin{eqnarray}
h(\eta) &=& I_1(\eta) + I_2(\eta) + e^{-\pi\eta} I_3(\eta) ,
  \quad {\rm where} \\
I_1(\eta) &=& \int_0^\pi \left[\frac{1}{t}-
  \frac{1}{2\tan(t/2)}\right] e^{-\eta t} \, dt ,\\
I_2(\eta) &=& \int_\pi^\infty \frac{e^{-\eta t}}{t} \, dt , {\quad} {\rm  and}\\
I_3(\eta) &=& \int_0^\pi \frac{1}{2\tan(t/2)} \,
  \frac{\sinh(\eta t)}{\sinh(\eta\pi)} \, dt
\end{eqnarray}
\end{subequations}
that is sufficient to demonstrate that 
\begin{equation}
h(\eta)>0 \quad {\rm and} \quad \frac{dh}{d\eta} <0
\label{eq:h_inequality}
\end{equation}
for the repulsive Coulomb field.
These results are useful for determining the sign of the energy
derivative of $\sigma_\ell$ and/or $h(\eta)$ in this work.
To the best of our knowledge, these results regarding the monotonic
properties of $\sigma_\ell$ and $h(\eta)$ have not been noted previously.
Finally, we note that for $\eta\rightarrow\infty$
we have asymptotically~\cite[Eq.~6.3.19, p.~259]{Abr65}
\begin{equation}
h(\eta) \sim \frac{1}{12\eta^2} + \frac{1}{120\eta^4} + \ldots,
\label{eq:h_asymp}
\end{equation}
which is consistent with the Eq.~(\ref{eq:h_inequality}).

\section{Limiting forms for small and large \texorpdfstring{$\bm{\rho}$}{rho}}
\label{app:coulomb_rho}

We present in Table~\ref{tab:coulomb} the leading behavior of the various
Coulomb quantities used in this work for $\rho\rightarrow 0$ and
$\rho\rightarrow\infty$, considering $\eta$ to be constant.
The small-$\rho$ forms are deduced starting from Eqs.~14.1.3-14.1.23
of \textcite[pp.~539-540]{Abr65}, with the Gamow factor defined to be
\begin{equation}
C_\ell(\eta) =
  \frac{ 2^\ell e^{-\pi\eta/2}
  \left[ \Gamma(\ell+1+i\eta)\Gamma(\ell+1-i\eta)\right]^{1/2} }
  {\Gamma(2\ell+2)}.
\end{equation}
In order to extract $\partial S_\ell/\partial E$ as $\rho$
(or the radius) goes to zero, it is necessary to consider the expansion
of $S_\ell$ beyond the leading term given in Table~\ref{tab:coulomb}.
For $\ell$ a non-negative integer, this results in
\begin{equation}
S_\ell = \left\{ \begin{array}{ll} 
 {\displaystyle 2\eta\rho \left[ \log(2\eta\rho) +
   2\gamma + h(\eta) \right] +\ldots } &
   \ell=0 \\
 {\displaystyle -\ell -\frac{\eta\rho}{\ell} +
   \frac{1+(\eta/\ell)^2}{2\ell-1}\rho^2 +\ldots } &
   \ell>0
 \end{array} \right. ,
\label{eq:s_small_rho}
\end{equation}
where $\gamma$ and $h(\eta)$ are defined in Appendix~\ref{app:cphase_h}.
The resulting expressions for the energy dependence of the shift factor
in the $\rho\rightarrow 0$ limit (with $\eta$ fixed) are
\begin{equation}
\frac{\partial S_\ell}{\partial E} = \left\{ \begin{array}{ll}
  {\displaystyle -\frac{\eta^2\rho}{E}\frac{dh}{d\eta} +\ldots} &
    \quad\ell=0 \\[3mm]
  {\displaystyle \frac{2\mu r^2}{\hbar^2(2\ell-1)}+\ldots } & \quad\ell >0
  \end{array} \right. .
\label{eq:dsde_small_rho}
\end{equation}
As discussed in Appendix~\ref{app:cphase_h}, $dh/d\eta<0$,
and consequently $\partial S_\ell/\partial E>0$ in this limit.
It is also interesting to note that for $\ell>0$ the quantity
$\partial S_\ell/\partial E$ is independent of the energy and Coulomb field
for sufficiently small radii.

The large-$\rho$ forms are deduced from asymptotic expansions
given by Eqs.~14.5.1-14.5.9 of Ref.~\cite[pp.~539-540]{Abr65}.
Energy derivatives of some the Coulomb quantities were determined
for the large-$\rho$ limit and are given in Table~\ref{tab:asymp_de}.
They are useful for evaluating the surface terms of integrals that arise
in this work.
In addition, one can see that $\partial S_\ell/\partial E>0$ in the
large-$\rho$ limit.

\begin{table}[tbh]
\caption{Energy derivatives of some Coulomb quantities for
large $\rho$.}
\label{tab:asymp_de}
\begin{ruledtabular}
\begin{tabular}{l}
$\frac{\partial A_\ell^2}{\partial E} \sim \frac{\rho}{2E}
  \left[ -\frac{2\eta}{\rho^2}
  -\frac{6\eta^2+\ell(\ell+1)}{\rho^3} +\ldots \right]$ \\[2ex]
$\frac{\partial\phi_\ell}{\partial E} \sim \frac{\rho}{2E}
  \left[ 1 + \frac{\eta}{\rho}\log(2\rho) -\frac{\eta}{\rho}
  -\frac{3\eta^2+\ell(\ell+1)}{2\rho^2} + \ldots \right]
  + \frac{\partial\sigma_\ell}{\partial E}$ \\[2ex]
$\frac{\partial P_\ell}{\partial E} \sim \frac{\rho}{2E}
  \left[ 1 + \frac{\eta}{\rho} + \frac{3\eta^2+\ell(\ell+1)}{2\rho^2}
  + \ldots \right]$ \\[2ex]
$\frac{\partial S_\ell}{\partial E} \sim \frac{\rho}{2E}
  \left[ \frac{\eta}{\rho^2} +
  \frac{4\eta^2+\ell(\ell+1)}{\rho^3} + \ldots \right]$
\end{tabular}
\end{ruledtabular}
\end{table}

\section{Limiting forms for small and large \textit{E}}
\label{app:coulomb_e}

We will first consider the case of large energies at fixed radius, where
$\rho\rightarrow\infty$ and $\eta\rightarrow 0$.
Since the asymptotic formulas for large $\rho$ are still valid when
$\eta$ is small, the high-energy limits can be calculated using
the $\rho\rightarrow\infty$ formulas located in Tables~\ref{tab:coulomb}
and~\ref{tab:asymp_de} of Appendix~\ref{app:coulomb_rho}.

For considering the low-energy limit at fixed radius, the
expansions in terms of modified Bessel functions
and inverse powers of $\eta^2$ are appropriate~\cite{Abr54,Hum85}. 
We will focus our attention on the shift factor, as the
penetration factor has been thoroughly covered elsewhere~\cite{Lan58,Hum87}.
In this limit, it is convenient to use the energy-independent radial
coordinate $x_0$:
\begin{equation}
x_0=\sqrt{8\eta\rho}=\sqrt{8\alpha r},
\end{equation}
where $\alpha=\eta k$ is likewise independent of energy. In the low-energy
limit, we have $G_\ell \gg F_\ell$ and the shift factor at zero energy is
given by the well-known result~\cite{Lan58,Pro58}
\begin{equation}
S_\ell = -\ell -\frac{x_0 K_{2\ell}(x_0)}{2 K_{2\ell+1}(x_0)},
\end{equation}
where $K_\nu$ are the irregular modified Bessel functions.
The energy derivative of the shift factor can be found by considering
the leading energy-dependent terms in the expansions of $G_\ell$ and $G_\ell'$.
The slope of the shift factor at zero energy is thus found:
\begin{align} \label{eq:dsde_zero1}
\frac{\partial S_\ell}{\partial E} &= \frac{2\mu}{\hbar^2\alpha^2}
  \frac{x_0^3}{192 [K_{2\ell+1}(x_0)]^2} \\
  &\times \biggl\{ 6(\ell+1)\left[
  K_{2\ell+1}(x_0)K_{2\ell+2}(x_0) \right. \nonumber \\
  & \left. \quad -K_{2\ell}(x_0)K_{2\ell+3}(x_0)\right]
  +x_0\left[ K_{2\ell}(x_0)K_{2\ell+4}(x_0) \right. \nonumber \\
  & \left. \quad -K_{2\ell+1}(x_0)K_{2\ell+3}(x_0)
  \right] \biggr\}. \nonumber
\end{align}
\textcite[p.~351]{Lan58} have given in their Eq.~(A.25) a similar expression,
valid for $\ell=0$ only, that is equivalent to our result in that case.
It is not at all clear that $\partial S_\ell/\partial E$ as
given by Eq.~(\ref{eq:dsde_zero1}) is positive.
An alternative approach is to realize that at zero energy the wave function
decays exponentially at large radii all the way out to $\infty$ (physically,
the classical turning radius is infinite) and
Eq.~(\ref{eq:dlde_a}) can be used. This results in
\begin{equation}
\frac{\partial S_\ell}{\partial E} = \frac{2\mu}{\hbar^2\alpha^2} \,
  \frac{x_0^2}{32}\int_{x_0}^\infty 
  \left[ \frac{x K_{2\ell+1}(x)}{x_0 K_{2\ell+1}(x_0)} \right]^2 x \, dx ,
\label{eq:dsde_zero2}
\end{equation}
which clearly shows $\partial S_\ell/\partial E >0$ at zero energy.
The equivalence of Eqs.~(\ref{eq:dsde_zero1}) and~(\ref{eq:dsde_zero2})
can be confirmed using differentiation and recurrence formulas.
The small radius ($x_0\rightarrow 0$) limits of these results,
for $\ell$ a non-negative integer, are:
\begin{equation}
S_\ell = \left\{ \begin{array}{ll}
  \displaystyle{\frac{x_0^2}{2}\left[ \gamma+\log(x_0/2)\right] +\ldots} &
  \ell=0 \\
  \displaystyle{ -\ell-\frac{x_0^2}{8\ell} +\ldots } &
  \ell>0 \end{array}\right.
  \label{eq:shift_zero_e} \\[3mm]
\end{equation}
and
\begin{equation}
\frac{\partial S_\ell}{\partial E} = \frac{2\mu}{\hbar^2\alpha^2}
  \left\{ \begin{array}{ll}
  \displaystyle{ \frac{x_0^2}{48}+\ldots } & \ell=0 \\
  \displaystyle{ \frac{x_0^4}{64(2\ell-1) } +\ldots }&
  \ell>0 \end{array}\right. ,
\end{equation}
which are consistent with Eqs.~(\ref{eq:s_small_rho})
and~(\ref{eq:dsde_small_rho}) when the low-energy behavior of
$h(\eta)$ is taken into consideration via Eq.~(\ref{eq:h_asymp}).

\section{Additional integral relations}
\label{app:integrals}

Some integral expressions involving $\partial S/\partial E$ are given here.
A general class of relations may be derived by multiplying
through by an arbitrary function $f$ before integrating to achieve
Eq.~(\ref{eq:int_o1o2}). This procedure results in:
\begin{equation}
-E\left[\frac{fO^2}{\rho}\frac{\partial L}{\partial E}
  \right]_{\rho_a}^{\rho_b} = \int_{\rho_a}^{\rho_b} O^2\left[
  f-\frac{E}{\rho}f'\frac{\partial L}{\partial E} \right] \, d\rho .
\label{eq:o2f}
\end{equation}
One choice for $f$ is
\begin{equation}
f=e^{-i\psi},
\end{equation}
where $\psi$ is defined by Eq.~(\ref{eq:def_psi}),
such that $fO^2=A^2$, $f'=-2if/A^2$, and
\begin{equation}
-E\left[\frac{A^2}{\rho}\frac{\partial L}{\partial E} \right]_{\rho_a}^{\rho_b}
  = \int_{\rho_a}^{\rho_b} A^2 \left[1+ \frac{2i}{A^2}\frac{E}{\rho}
  \frac{\partial L}{\partial E} \right] \, d\rho.
\end{equation}
Taking the real part gives
\begin{equation}
-E\left[ \frac{A^2}{\rho} \frac{\partial S}{\partial E}\right]_{\rho_a}^{\rho_b}
  = \int_{\rho_a}^{\rho_b} \left[ A^2 -\frac{2E}{\rho}\frac{\partial P}{\partial E}
  \right] \, d\rho .
\label{eq:int_rel_ab}
\end{equation}
and then letting $\rho_b\rightarrow\infty$ (noting that the integrand
${\sim 1/\rho^3}$ for large $\rho$) yields a relation for
${\partial S}/{\partial E}$:
\begin{equation}
E\left[ \frac{A^2}{\rho} \frac{\partial S}{\partial E}\right]_{\rho_a} =
  \int_{\rho_a}^\infty \left[ A^2 -\frac{2E}{\rho}\frac{\partial P}{\partial E}
  \right] \, d\rho .
\end{equation}
Interestingly, Eq.~(\ref{eq:p_upper}) ensures that the integrand
in the r.h.s. of this equation is positive.
Alternatively, noting that
\begin{equation}
\frac{d\phi}{dr}=\frac{P}{r} \quad {\rm and~hence} \quad
  \frac{\partial}{\partial E}\left( \frac{d\phi}{dr}\right)=
  \frac{1}{r}\frac{\partial P}{\partial E},
\end{equation}
the second term in the integrand of Eq.~(\ref{eq:int_rel_ab}) can be
integrated to give
\begin{equation}
E\left[ -\frac{A^2}{\rho} \frac{\partial S}{\partial E} +
  2\frac{\partial\phi}{\partial E} \right]_{\rho_a}^{\rho_b} =
  \int_{\rho_a}^{\rho_b} A^2 \, d\rho ,
\end{equation}
which happens to be equivalent to Eq.~(A.31) of
Lane and Thomas~\cite[p.~352]{Lan58}.
The leading asymptotic behavior of $A^2$ for large $\rho$ can be subtracted
\begin{equation}
\begin{split}
\biggl[ -E\frac{A^2}{\rho}\frac{\partial S}{\partial E} +
  2E\frac{\partial\phi}{\partial E} & -\rho -\eta\log(2\rho)
  \biggr]_{\rho_a}^{\rho_b} \\ &= \int_{\rho_a}^{\rho_b}
  (A^2-1-\frac{\eta}{\rho}) \, d\rho 
\end{split}
\end{equation}
in order to allow $\rho_b\rightarrow\infty$ to be taken:
\begin{equation}
\begin{split}
\biggl[ E\frac{A^2}{\rho}&\frac{\partial S}{\partial E} -
  2E\left(\frac{\partial\phi}{\partial E}-
  \frac{\partial\sigma}{\partial E}\right)
  +\rho \\ &+\eta\log(2\rho) + \eta \biggr]_{\rho_a} =
  \int_{\rho_a}^{\infty} (A^2-1-\frac{\eta}{\rho}) \, d\rho ,
\end{split}
\end{equation}
where the asymptotic forms of the functions have been used to evaluate
the surface terms at $\infty$.

It is also natural to investigate the
integral relations arising from considering solutions
$O_1^*$ and $O_2$ with $O_2\rightarrow O_1$.
Assuming that $O=A\exp(i\phi)$ and multiplying through by an
arbitrary function $g$ before integrating yields:
\begin{equation}
\begin{split}
-E\biggl[&g \left( \frac{A^2}{\rho}\frac{\partial S}{\partial E}
  -2\frac{\partial\phi}{\partial E} \right) \biggr]_{\rho_a}^{\rho_b} \\
&= \int_{\rho_a}^{\rho_b} \left[ gA^2 - Eg' \left(
  \frac{A^2}{\rho}\frac{\partial S}{\partial E}
  -2\frac{\partial\phi}{\partial E} \right) \right] \, d\rho .
\end{split}
\end{equation}
These relations are not independent of those derivable from
Eq.~(\ref{eq:o2f}).

\bibliography{coulomb.bib}

\begin{thebibliography}{27}%
\makeatletter
\providecommand \@ifxundefined [1]{%
 \@ifx{#1\undefined}
}%
\providecommand \@ifnum [1]{%
 \ifnum #1\expandafter \@firstoftwo
 \else \expandafter \@secondoftwo
 \fi
}%
\providecommand \@ifx [1]{%
 \ifx #1\expandafter \@firstoftwo
 \else \expandafter \@secondoftwo
 \fi
}%
\providecommand \natexlab [1]{#1}%
\providecommand \enquote  [1]{``#1''}%
\providecommand \bibnamefont  [1]{#1}%
\providecommand \bibfnamefont [1]{#1}%
\providecommand \citenamefont [1]{#1}%
\providecommand \href@noop [0]{\@secondoftwo}%
\providecommand \href [0]{\begingroup \@sanitize@url \@href}%
\providecommand \@href[1]{\@@startlink{#1}\@@href}%
\providecommand \@@href[1]{\endgroup#1\@@endlink}%
\providecommand \@sanitize@url [0]{\catcode `\\12\catcode `\$12\catcode
  `\&12\catcode `\#12\catcode `\^12\catcode `\_12\catcode `\%12\relax}%
\providecommand \@@startlink[1]{}%
\providecommand \@@endlink[0]{}%
\providecommand \url  [0]{\begingroup\@sanitize@url \@url }%
\providecommand \@url [1]{\endgroup\@href {#1}{\urlprefix }}%
\providecommand \urlprefix  [0]{URL }%
\providecommand \Eprint [0]{\href }%
\providecommand \doibase [0]{http://dx.doi.org/}%
\providecommand \selectlanguage [0]{\@gobble}%
\providecommand \bibinfo  [0]{\@secondoftwo}%
\providecommand \bibfield  [0]{\@secondoftwo}%
\providecommand \translation [1]{[#1]}%
\providecommand \BibitemOpen [0]{}%
\providecommand \bibitemStop [0]{}%
\providecommand \bibitemNoStop [0]{.\EOS\space}%
\providecommand \EOS [0]{\spacefactor3000\relax}%
\providecommand \BibitemShut  [1]{\csname bibitem#1\endcsname}%
\let\auto@bib@innerbib\@empty
\bibitem [{\citenamefont {Lane}\ and\ \citenamefont {Thomas}(1958)}]{Lan58}%
  \BibitemOpen
  \bibfield  {author} {\bibinfo {author} {\bibfnamefont {A.~M.}\ \bibnamefont
  {Lane}}\ and\ \bibinfo {author} {\bibfnamefont {R.~G.}\ \bibnamefont
  {Thomas}},\ }\bibfield  {title} {\enquote {\bibinfo {title}
  {\textit{R}-matrix theory of nuclear reactions},}\ }\href {\doibase
  10.1103/RevModPhys.30.257} {\bibfield  {journal} {\bibinfo  {journal} {Rev.
  Mod. Phys.}\ }\textbf {\bibinfo {volume} {30}},\ \bibinfo {pages} {257--353}
  (\bibinfo {year} {1958})}\BibitemShut {NoStop}%
\bibitem [{\citenamefont {Descouvemont}\ and\ \citenamefont
  {Baye}(2010)}]{Des10}%
  \BibitemOpen
  \bibfield  {author} {\bibinfo {author} {\bibfnamefont {P.}~\bibnamefont
  {Descouvemont}}\ and\ \bibinfo {author} {\bibfnamefont {D.}~\bibnamefont
  {Baye}},\ }\bibfield  {title} {\enquote {\bibinfo {title} {The
  \textit{R}-matrix theory},}\ }\href {\doibase 10.1088/0034-4885/73/3/036301}
  {\bibfield  {journal} {\bibinfo  {journal} {Reports on Progress in Physics}\
  }\textbf {\bibinfo {volume} {73}},\ \bibinfo {pages} {036301} (\bibinfo
  {year} {2010})}\BibitemShut {NoStop}%
\bibitem [{\citenamefont {Brune}(2002)}]{Bru02}%
  \BibitemOpen
  \bibfield  {author} {\bibinfo {author} {\bibfnamefont {C.~R.}\ \bibnamefont
  {Brune}},\ }\bibfield  {title} {\enquote {\bibinfo {title} {Alternative
  parametrization of \textit{R}-matrix theory},}\ }\href {\doibase
  10.1103/PhysRevC.66.044611} {\bibfield  {journal} {\bibinfo  {journal} {Phys.
  Rev. C}\ }\textbf {\bibinfo {volume} {66}},\ \bibinfo {pages} {044611}
  (\bibinfo {year} {2002})}\BibitemShut {NoStop}%
\bibitem [{\citenamefont {Wheeler}(1937)}]{Whe37}%
  \BibitemOpen
  \bibfield  {author} {\bibinfo {author} {\bibfnamefont {John~A.}\ \bibnamefont
  {Wheeler}},\ }\bibfield  {title} {\enquote {\bibinfo {title} {Wave functions
  for large arguments by the amplitude-phase method},}\ }\href {\doibase
  10.1103/PhysRev.52.1123} {\bibfield  {journal} {\bibinfo  {journal} {Phys.
  Rev.}\ }\textbf {\bibinfo {volume} {52}},\ \bibinfo {pages} {1123--1127}
  (\bibinfo {year} {1937})}\BibitemShut {NoStop}%
\bibitem [{\citenamefont {Thaler}\ and\ \citenamefont
  {Biedenharn}(1957)}]{Tha57}%
  \BibitemOpen
  \bibfield  {author} {\bibinfo {author} {\bibfnamefont {R.~M.}\ \bibnamefont
  {Thaler}}\ and\ \bibinfo {author} {\bibfnamefont {L.~C.}\ \bibnamefont
  {Biedenharn}},\ }\bibfield  {title} {\enquote {\bibinfo {title} {A note on
  the calculation of {C}oulomb penetration and {C}oulomb wave functions},}\
  }\href {\doibase 10.1016/0029-5582(57)90107-4} {\bibfield  {journal}
  {\bibinfo  {journal} {Nuclear Physics}\ }\textbf {\bibinfo {volume} {3}},\
  \bibinfo {pages} {207--212} (\bibinfo {year} {1957})}\BibitemShut {NoStop}%
\bibitem [{\citenamefont {Prosser}\ and\ \citenamefont
  {Biedenharn}(1958)}]{Pro58}%
  \BibitemOpen
  \bibfield  {author} {\bibinfo {author} {\bibfnamefont {F.~W.}\ \bibnamefont
  {Prosser}}\ and\ \bibinfo {author} {\bibfnamefont {L.~C.}\ \bibnamefont
  {Biedenharn}},\ }\bibfield  {title} {\enquote {\bibinfo {title} {Some
  properties of the shift and penetration factors in nuclear reactions},}\
  }\href {\doibase 10.1103/PhysRev.109.413} {\bibfield  {journal} {\bibinfo
  {journal} {Phys. Rev.}\ }\textbf {\bibinfo {volume} {109}},\ \bibinfo {pages}
  {413--417} (\bibinfo {year} {1958})}\BibitemShut {NoStop}%
\bibitem [{\citenamefont {Hull}\ and\ \citenamefont {Breit}(1959)}]{Hul59}%
  \BibitemOpen
  \bibfield  {author} {\bibinfo {author} {\bibfnamefont {M.~H.}\ \bibnamefont
  {Hull}, \bibfnamefont {Jr.}}\ and\ \bibinfo {author} {\bibfnamefont
  {G.}~\bibnamefont {Breit}},\ }\enquote {\bibinfo {title} {Coulomb wave
  functions},}\ in\ \href {\doibase 10.1007/978-3-642-45923-8_2} {\emph
  {\bibinfo {booktitle} {Nuclear Reactions II: Theory / Kernreaktionen II:
  Theorie}}}\ (\bibinfo  {publisher} {Springer},\ \bibinfo {address} {Berlin,
  Heidelberg},\ \bibinfo {year} {1959})\ pp.\ \bibinfo {pages}
  {408--465}\BibitemShut {NoStop}%
\bibitem [{\citenamefont {Erd{\' e}lyi}(1938)}]{Erd38}%
  \BibitemOpen
  \bibfield  {author} {\bibinfo {author} {\bibfnamefont {A.}~\bibnamefont
  {Erd{\' e}lyi}},\ }\bibfield  {title} {\enquote {\bibinfo {title} {{LXXV}.
  {I}ntegral representations for products of {W}hittaker functions},}\ }\href
  {\doibase 10.1080/14786443808562178} {\bibfield  {journal} {\bibinfo
  {journal} {The London, Edinburgh, and Dublin Philosophical Magazine and
  Journal of Science}\ }\textbf {\bibinfo {volume} {26}},\ \bibinfo {pages}
  {871--877} (\bibinfo {year} {1938})}\BibitemShut {NoStop}%
\bibitem [{\citenamefont {Buchholz}(1969)}]{Buc69}%
  \BibitemOpen
  \bibfield  {author} {\bibinfo {author} {\bibfnamefont {Herbert}\ \bibnamefont
  {Buchholz}},\ }\href {\doibase 10.1007/978-3-642-88396-5} {\emph {\bibinfo
  {title} {The Confluent Hypergeometric Function}}},\ Springer Tracts in
  Natural Philosophy, Volume 15\ (\bibinfo  {publisher} {Springer-Verlag},\
  \bibinfo {address} {Berlin},\ \bibinfo {year} {1969})\BibitemShut {NoStop}%
\bibitem [{\citenamefont {Miller}\ and\ \citenamefont {Samko}(2001)}]{Mil01}%
  \BibitemOpen
  \bibfield  {author} {\bibinfo {author} {\bibfnamefont {K.~S.}\ \bibnamefont
  {Miller}}\ and\ \bibinfo {author} {\bibfnamefont {S.~G.}\ \bibnamefont
  {Samko}},\ }\bibfield  {title} {\enquote {\bibinfo {title} {Completely
  monotonic functions},}\ }\href {\doibase 10.1080/10652460108819360}
  {\bibfield  {journal} {\bibinfo  {journal} {Integral Transforms and Special
  Functions}\ }\textbf {\bibinfo {volume} {12}},\ \bibinfo {pages} {389--402}
  (\bibinfo {year} {2001})}\BibitemShut {NoStop}%
\bibitem [{\citenamefont {Watson}(1944)}]{Wat44}%
  \BibitemOpen
  \bibfield  {author} {\bibinfo {author} {\bibfnamefont {G.~N.}\ \bibnamefont
  {Watson}},\ }\href@noop {} {\emph {\bibinfo {title} {A Treatise on the Theory
  of Bessel Functions}}},\ \bibinfo {edition} {2nd}\ ed.\ (\bibinfo
  {publisher} {Cambridge University Press},\ \bibinfo {address} {Cambridge},\
  \bibinfo {year} {1944})\BibitemShut {NoStop}%
\bibitem [{\citenamefont {Abramowitz}\ and\ \citenamefont
  {Stegun}(1965)}]{Abr65}%
  \BibitemOpen
  \bibfield  {author} {\bibinfo {author} {\bibfnamefont {M.}~\bibnamefont
  {Abramowitz}}\ and\ \bibinfo {author} {\bibfnamefont {I.}~\bibnamefont
  {Stegun}},\ }\href@noop {} {\emph {\bibinfo {title} {{H}andbook of
  {M}athematical {F}unctions}}}\ (\bibinfo  {publisher} {Dover Publications},\
  \bibinfo {address} {New York},\ \bibinfo {year} {1965})\BibitemShut {NoStop}%
\bibitem [{\citenamefont {Hartman}(1973)}]{Har73}%
  \BibitemOpen
  \bibfield  {author} {\bibinfo {author} {\bibfnamefont {Philip}\ \bibnamefont
  {Hartman}},\ }\bibfield  {title} {\enquote {\bibinfo {title} {On differential
  equations, {V}olterra equations and the function ${J}_\mu^2+{Y}_\mu^2$},}\
  }\href {\doibase 10.2307/2373730} {\bibfield  {journal} {\bibinfo  {journal}
  {American Journal of Mathematics}\ }\textbf {\bibinfo {volume} {95}},\
  \bibinfo {pages} {553--593} (\bibinfo {year} {1973})}\BibitemShut {NoStop}%
\bibitem [{\citenamefont {Lorch}\ and\ \citenamefont {Szego}(1963)}]{Lor63}%
  \BibitemOpen
  \bibfield  {author} {\bibinfo {author} {\bibfnamefont {Lee}\ \bibnamefont
  {Lorch}}\ and\ \bibinfo {author} {\bibfnamefont {Peter}\ \bibnamefont
  {Szego}},\ }\bibfield  {title} {\enquote {\bibinfo {title} {Higher
  monotonicity properties of certain {S}turm-{L}iouville functions},}\ }\href
  {\doibase 10.1007/BF02391809} {\bibfield  {journal} {\bibinfo  {journal}
  {Acta Math.}\ }\textbf {\bibinfo {volume} {109}},\ \bibinfo {pages} {55--73}
  (\bibinfo {year} {1963})}\BibitemShut {NoStop}%
\bibitem [{\citenamefont {Durand}(1978)}]{Dur78}%
  \BibitemOpen
  \bibfield  {author} {\bibinfo {author} {\bibfnamefont {Loyal}\ \bibnamefont
  {Durand}},\ }\bibfield  {title} {\enquote {\bibinfo {title} {Product formulas
  and {N}icholson-type integrals for {J}acobi functions. {I}: {S}ummary of
  results},}\ }\href {\doibase 10.1137/0509007} {\bibfield  {journal} {\bibinfo
   {journal} {SIAM Journal on Mathematical Analysis}\ }\textbf {\bibinfo
  {volume} {9}},\ \bibinfo {pages} {76--86} (\bibinfo {year}
  {1978})}\BibitemShut {NoStop}%
\bibitem [{\citenamefont {Muldoon}(1986)}]{Mul86}%
  \BibitemOpen
  \bibfield  {author} {\bibinfo {author} {\bibfnamefont {M.~E.}\ \bibnamefont
  {Muldoon}},\ }\enquote {\bibinfo {title} {On the zeros of some special
  functions: Differential equations and {N}icholson-type formulas},}\ in\ \href
  {\doibase 10.1007/BFb0076063} {\emph {\bibinfo {booktitle} {Equadiff 6:
  Proceedings of the International Conference on Differential Equations and
  their Applications held in Brno, Czechoslovakia, Aug. 26--30, 1985}}},\
  \bibinfo {editor} {edited by\ \bibinfo {editor} {\bibfnamefont {Jarom{\'i}r}\
  \bibnamefont {Vosmansk{\'y}}}\ and\ \bibinfo {editor} {\bibfnamefont
  {Milo{\v{s}}}\ \bibnamefont {Zl{\'a}mal}}}\ (\bibinfo  {publisher} {Springer
  Berlin Heidelberg},\ \bibinfo {address} {Berlin, Heidelberg},\ \bibinfo
  {year} {1986})\ pp.\ \bibinfo {pages} {155--160}\BibitemShut {NoStop}%
\bibitem [{\citenamefont {Hartman}(1961)}]{Har61}%
  \BibitemOpen
  \bibfield  {author} {\bibinfo {author} {\bibfnamefont {Philip}\ \bibnamefont
  {Hartman}},\ }\bibfield  {title} {\enquote {\bibinfo {title} {On differential
  equations and the function ${J}_\mu^2+{Y}_\mu^2$},}\ }\href {\doibase
  10.2307/2372726} {\bibfield  {journal} {\bibinfo  {journal} {American Journal
  of Mathematics}\ }\textbf {\bibinfo {volume} {83}},\ \bibinfo {pages}
  {154--188} (\bibinfo {year} {1961})}\BibitemShut {NoStop}%
\bibitem [{\citenamefont {Goldstein}\ and\ \citenamefont
  {Thaler}(1958)}]{Gol58}%
  \BibitemOpen
  \bibfield  {author} {\bibinfo {author} {\bibfnamefont {M.}~\bibnamefont
  {Goldstein}}\ and\ \bibinfo {author} {\bibfnamefont {R.~M.}\ \bibnamefont
  {Thaler}},\ }\bibfield  {title} {\enquote {\bibinfo {title} {Bessel functions
  for large arguments},}\ }\href {\doibase 10.1090/S0025-5718-1958-0102906-3}
  {\bibfield  {journal} {\bibinfo  {journal} {Math. Comp.}\ }\textbf {\bibinfo
  {volume} {12}},\ \bibinfo {pages} {18--26} (\bibinfo {year}
  {1958})}\BibitemShut {NoStop}%
\bibitem [{\citenamefont {Thomas}(1951)}]{Tho51}%
  \BibitemOpen
  \bibfield  {author} {\bibinfo {author} {\bibfnamefont {R.~G.}\ \bibnamefont
  {Thomas}},\ }\bibfield  {title} {\enquote {\bibinfo {title} {On the
  determination of reduced widths from the one-level dispersion formula},}\
  }\href {\doibase 10.1103/PhysRev.81.148} {\bibfield  {journal} {\bibinfo
  {journal} {Phys. Rev.}\ }\textbf {\bibinfo {volume} {81}},\ \bibinfo {pages}
  {148--149} (\bibinfo {year} {1951})}\BibitemShut {NoStop}%
\bibitem [{\citenamefont {Hale}\ \emph {et~al.}(1987)\citenamefont {Hale},
  \citenamefont {Brown},\ and\ \citenamefont {Jarmie}}]{Hal87}%
  \BibitemOpen
  \bibfield  {author} {\bibinfo {author} {\bibfnamefont {G.~M.}\ \bibnamefont
  {Hale}}, \bibinfo {author} {\bibfnamefont {Ronald~E.}\ \bibnamefont {Brown}},
  \ and\ \bibinfo {author} {\bibfnamefont {Nelson}\ \bibnamefont {Jarmie}},\
  }\bibfield  {title} {\enquote {\bibinfo {title} {Pole structure of the
  ${J}^\pi={3/2}^+$ resonance in $^{5}\mathrm{He}$},}\ }\href {\doibase
  10.1103/PhysRevLett.59.763} {\bibfield  {journal} {\bibinfo  {journal} {Phys.
  Rev. Lett.}\ }\textbf {\bibinfo {volume} {59}},\ \bibinfo {pages} {763--766}
  (\bibinfo {year} {1987})}\BibitemShut {NoStop}%
\bibitem [{\citenamefont {Tuck}(2006)}]{Tuc06}%
  \BibitemOpen
  \bibfield  {author} {\bibinfo {author} {\bibfnamefont {E.O.}\ \bibnamefont
  {Tuck}},\ }\bibfield  {title} {\enquote {\bibinfo {title} {On positivity of
  {F}ourier transforms},}\ }\href {\doibase 10.1017/S0004972700047511}
  {\bibfield  {journal} {\bibinfo  {journal} {Bulletin of the Australian
  Mathematical Society}\ }\textbf {\bibinfo {volume} {74}},\ \bibinfo {pages}
  {133--138} (\bibinfo {year} {2006})}\BibitemShut {NoStop}%
\bibitem [{\citenamefont {Bethe}(1949)}]{Bet49}%
  \BibitemOpen
  \bibfield  {author} {\bibinfo {author} {\bibfnamefont {H.~A.}\ \bibnamefont
  {Bethe}},\ }\bibfield  {title} {\enquote {\bibinfo {title} {Theory of the
  effective range in nuclear scattering},}\ }\href {\doibase
  10.1103/PhysRev.76.38} {\bibfield  {journal} {\bibinfo  {journal} {Phys.
  Rev.}\ }\textbf {\bibinfo {volume} {76}},\ \bibinfo {pages} {38--50}
  (\bibinfo {year} {1949})}\BibitemShut {NoStop}%
\bibitem [{\citenamefont {Jackson}\ and\ \citenamefont {Blatt}(1950)}]{Jac50}%
  \BibitemOpen
  \bibfield  {author} {\bibinfo {author} {\bibfnamefont {J.~David}\
  \bibnamefont {Jackson}}\ and\ \bibinfo {author} {\bibfnamefont {John~M.}\
  \bibnamefont {Blatt}},\ }\bibfield  {title} {\enquote {\bibinfo {title} {The
  interpretation of low energy proton-proton scattering},}\ }\href {\doibase
  10.1103/RevModPhys.22.77} {\bibfield  {journal} {\bibinfo  {journal} {Rev.
  Mod. Phys.}\ }\textbf {\bibinfo {volume} {22}},\ \bibinfo {pages} {77--118}
  (\bibinfo {year} {1950})}\BibitemShut {NoStop}%
\bibitem [{\citenamefont {van Haeringen}(1977)}]{Hae77}%
  \BibitemOpen
  \bibfield  {author} {\bibinfo {author} {\bibfnamefont {H.}~\bibnamefont {van
  Haeringen}},\ }\bibfield  {title} {\enquote {\bibinfo {title} {\textit{T}
  matrix and effective range function for {C}oulomb plus rational separable
  potentials especially for \textit{l}=1},}\ }\href {\doibase 10.1063/1.523373}
  {\bibfield  {journal} {\bibinfo  {journal} {Journal of Mathematical Physics}\
  }\textbf {\bibinfo {volume} {18}},\ \bibinfo {pages} {927--940} (\bibinfo
  {year} {1977})}\BibitemShut {NoStop}%
\bibitem [{\citenamefont {Abramowitz}(1954)}]{Abr54}%
  \BibitemOpen
  \bibfield  {author} {\bibinfo {author} {\bibfnamefont {Milton}\ \bibnamefont
  {Abramowitz}},\ }\bibfield  {title} {\enquote {\bibinfo {title} {Regular and
  irregular {C}oulomb wave functions expressed in terms of {B}essel-{C}lifford
  functions},}\ }\href {\doibase 10.1002/sapm1954331111} {\bibfield  {journal}
  {\bibinfo  {journal} {Journal of Mathematics and Physics}\ }\textbf {\bibinfo
  {volume} {33}},\ \bibinfo {pages} {111--116} (\bibinfo {year}
  {1954})}\BibitemShut {NoStop}%
\bibitem [{\citenamefont {Humblet}(1985)}]{Hum85}%
  \BibitemOpen
  \bibfield  {author} {\bibinfo {author} {\bibfnamefont {J.}~\bibnamefont
  {Humblet}},\ }\bibfield  {title} {\enquote {\bibinfo {title} {Bessel function
  expansions of {C}oulomb wave functions},}\ }\href {\doibase 10.1063/1.526602}
  {\bibfield  {journal} {\bibinfo  {journal} {Journal of Mathematical Physics}\
  }\textbf {\bibinfo {volume} {26}},\ \bibinfo {pages} {656--659} (\bibinfo
  {year} {1985})}\BibitemShut {NoStop}%
\bibitem [{\citenamefont {Humblet}\ \emph {et~al.}(1987)\citenamefont
  {Humblet}, \citenamefont {Fowler},\ and\ \citenamefont {Zimmerman}}]{Hum87}%
  \BibitemOpen
  \bibfield  {author} {\bibinfo {author} {\bibfnamefont {J.}~\bibnamefont
  {Humblet}}, \bibinfo {author} {\bibfnamefont {W.~A.}\ \bibnamefont {Fowler}},
  \ and\ \bibinfo {author} {\bibfnamefont {B.~A.}\ \bibnamefont {Zimmerman}},\
  }\bibfield  {title} {\enquote {\bibinfo {title} {Approximate penetration
  factors for nuclear reactions of astrophysical interest},}\ }\href
  {http://adsabs.harvard.edu/abs/1987A%26A...177..317H} {\bibfield  {journal}
  {\bibinfo  {journal} {Astronomy and Astrophysics}\ }\textbf {\bibinfo
  {volume} {177}},\ \bibinfo {pages} {317--325} (\bibinfo {year}
  {1987})}\BibitemShut {NoStop}%
\end{thebibliography}%

\end{document}